\documentclass{aastex631}

\usepackage{amsmath,amstext}
\usepackage[T1]{fontenc}
\usepackage[figure,figure*]{hypcap}
\usepackage{graphicx}
\usepackage{subfigure}
\usepackage{tabularx}
\usepackage{enumitem}
\usepackage{appendix}

\newcommand{\be}{\begin{equation}}
\newcommand{\ee}{\end{equation}}

\shorttitle{Late-Time Radio TDEs}
\shortauthors{Cendes et al.}

\graphicspath{{./}{figures/}}

\begin{document}

\title{Ubiquitous Late Radio Emission from Tidal Disruption Events}

\correspondingauthor{Yvette Cendes}
\email{yvette.cendes@cfa.harvard.edu}

\author[0000-0001-7007-6295]{Y. Cendes}
\affiliation{Center for Astrophysics | Harvard \& Smithsonian,
Cambridge, MA 02138, USA}
\affiliation{Department of Physics, University of Oregon, Eugene, OR 97403, USA}

\author[0000-0002-9392-9681]{E. Berger}
\affiliation{Center for Astrophysics | Harvard \& Smithsonian, Cambridge, MA 02138, USA}

\author[0000-0002-8297-2473]{K. D. Alexander}
\affiliation{Department of Astronomy and Steward Observatory, University of Arizona, 933 North Cherry Avenue, Tucson, AZ 85721-0065, USA}

\author[0000-0002-7706-5668]{R. Chornock}
\affiliation{Department of Astronomy, University of California, Berkeley, CA 94720-3411, USA}

\author[0000-0003-4768-7586]{R. Margutti}
\affiliation{Department of Astronomy, University of California, Berkeley, CA 94720-3411, USA}
\affiliation{Department of Physics, University of California, 366 Physics North MC 7300, Berkeley, CA 94720, USA}

\author[0000-0002-4670-7509]{B. Metzger}
\affiliation{Department of Physics and Columbia Astrophysics Laboratory, Columbia University, Pupin Hall, New York, NY 10027, USA}
\affiliation{Center for Computational Astrophysics, Flatiron Institute, 162 5th Ave, New York, NY 10010, USA}

\author[0000-0002-7721-8660]{M. H. Wieringa}
\affiliation{CSIRO Space and Astronomy,PO Box 76,Epping NSW 1710,Australia}

\author[0000-0002-0592-4152]{M. F. Bietenholz}
\affiliation{Department of Physics and Astronomy, York University, 4700 Keele St., Toronto, M3J~1P3, Ontario, Canada}

\author[0000-0003-2349-101X]{A. Hajela}
\affiliation{DARK, Niels Bohr Institute , University of Copenhagen, Jagtvej 128, DK-2200 Copenhagen, Denmark}

\author[0000-0003-1792-2338]{T. Laskar}
\affiliation{Department of Physics \& Astronomy, University of Utah, Salt Lake City, UT 84112, USA}
\affiliation{Department of Astrophysics/IMAPP, Radboud University, PO Box 9010, 6500 GL, The Netherlands}

\author{M. C. Stroh}
\affiliation{Center for Interdisciplinary Exploration and Research in Astrophysics (CIERA) and Department of Physics and Astronomy, Northwestern University, Evanston, IL 60208, USA}

\author{G. Terreran}
\affiliation{Las Cumbres Observatory, 6740 Cortona Drive, Suite 102, Goleta, CA 93117-5575, USA}

\begin{abstract}
We present radio observations of $23$ optically-discovered tidal disruption events (TDEs) on timescales of $\sim500-3200$ days post-discovery.  We detect 9 new TDEs that did not have detectable radio emission at earlier times, indicating a late-time brightening after several hundred (and up to 2300) days; an additional 7 TDEs exhibit radio emission whose origin is ambiguous or may be attributed to the host galaxy or an AGN. We also report a new rising component in one TDE previously detected in the radio at $\sim 10^3$ days. While the radio emission in some of the detected TDEs peaked on a timescale $\approx 2-4$ years, over half of the sample still shows rising emission. The range of luminosities for the sample is $\sim 10^{37}-10^{39}$ erg s$^{-1}$, about two orders of magnitude below the radio luminosity of the relativistic TDE Sw1644+57. Our data set indicates $\sim$40\% of all optical TDEs are detected in radio hundreds to thousands of days after discovery, and that this is probably more common than early radio emission peaking at $\sim 10^2$ days. Using an equipartition analysis, we find evidence for a delayed launch of the radio-emitting outflows, with delay timescales of $\sim 500-2000$ days, inferred velocities of $\approx 0.02-0.15c$, and kinetic energies of $\sim 10^{47}-10^{49}$ erg.  We rule out off-axis relativistic jets as a viable explanation for this population, and conclude delayed outflows are a more likely explanation, possibly from delayed disk formation. We conclude late radio emission marks a fairly ubiquitous but heretofore overlooked phase of TDE evolution.
\end{abstract}


\section{Introduction} 
\label{sec:intro}

Optical/UV and X-ray emissions from TDEs are generally thought to track the mass fallback (e.g., \citealt{Stone2013,Guillochon2013}) and cooling (e.g., \citealt{Metzger2022}) of the bound stellar debris onto the central supermassive black hole.  Radio observations, on the other hand, can reveal and characterize outflows from TDEs \citep{Alexander2020}, including the presence of relativistic jets \citep[e.g.~][]{Zauderer2011,Giannios11,DeColle2012,Andreoni2022}.  

To date, rapid follow-up of TDEs within the first days to weeks after discovery has led to the radio detection of several events.  These include most prominently Swift J1644+57 (Sw\,J1644+57), whose radio and mm emission are powered by a relativistic outflow with an energy of $\sim 10^{52}$ erg and an initial Lorentz factor of $\Gamma\sim 10$, first detected in radio $\sim2 $ days after its initial discovery in gamma-rays \citep{Zauderer2011,Metzger2012,p1,p2,Eftekhari2018,Cendes2021}. Other events, such as ASASSN-14li and AT2019dsg, have instead exhibited evidence for non-relativistic outflows, with $E_K\sim 10^{48}-10^{49}$ erg and $\beta\approx 0.05-0.1$, with first radio detections within a few weeks of discovery (e.g., \citealt{Alexander2020,Pasham2018,Alexander2016,Cendes2021b,Stein2021}).

Currently, most studies of radio TDEs primarily focus on the detection of radio emission at time in days $t_d < 100$ days, and emission is detected in $\sim20-30\%$ of cases \citep{Alexander2020}.  Recently, however, four TDEs have been reported to show radio emission with a delay relative to the time of optical discovery \footnote{IGRJ12580 is an X-ray discovered possible TDE with a delayed radio flare \citep{Perlman2022}, but is outside the scope of this paper which is focused on optically selected TDEs.}.  AT2018hyz was first detected at $\sim1000$ days post-discovery, despite several constraining upper limits at earlier times (including at 705 days), and has been increasing at a steep rise of $F_\nu\propto t^5$ relative to the time of optical discovery \citep{Cendes2022}.  This delay and rapid rise have been interpreted as either due to a delayed, mildly relativistic outflow launched $\approx 750$ d after optical discovery, or an off-axis jet which was launched promptly at the time of optical discovery and whose emitting area and angle have increased over time \citep{Cendes2022,Matsumoto2023}.  ASASSN-15oi was first detected $\approx 180$ days after optical discovery with a luminosity that exceeded earlier radio limits by a factor of $\approx 20$ \citep{Horesh2021}; this emission subsequently declined until about 550 days, and then exhibited a second rapid rise with a detection at 1400 days with an even higher luminosity than the first peak; see Figure~\ref{fig:lumin-tde}.  iPTF16fnl was first detected $\approx 150$ days after optical discovery, with a luminosity about a factor of 8 times larger than earlier limits (extending to 63 days) and appeared to slowly brighten to about 417 days \citep{Horesh2021b}.  The initial abrupt rise in ASASSN-15oi seems to differ from the radio light curve of AT2019dsg, although both reach their peak radio luminosity on a similar timescale and at a similar level.  The gradual rise and much lower peak luminosity of iPTF16fnl ($\approx 10^{37}$ erg s$^{-1}$), on the other hand, may indicate that it is simply a less energetic example of typical radio-emitting TDEs.  Finally, it is possible that TDEs with prompt detections can exhibit a secondary re-brightening after emission has faded: in the case of AT2020vwl, which was first detected in radio $\sim120$ days post-disruption, the source declined in emission to $\sim430$ days and is now increasing in brightness 900 days post-optical discovery \citep{Goodwin2023,Goodwin2023C}.

Here, we present radio observations of a sample of 23 optically detected TDEs on a timescale of about 500 to 2500 days post disruption, which show $\sim40\%$ of these TDEs have radio detections at these late times despite no emission detected at earlier times.  Our extensive multi-frequency data allow us to present detailed physical parameters for this TDE population in energetics, densities, and luminosities.  This more than doubles the number of radio-detected TDEs to date, and allow us to show an increasing diversity in the TDE landscape, particularly at these later time scales.

Our paper is structured as follows.  In \S\ref{sec:obs}, we present our radio observations.  In \S\ref{sec:results}, we discuss our results for individual TDEs in our study, including radio luminosity and evolution for individual TDEs and the rate of late-time TDE radio emission.  In \S\ref{sec:sed} we discuss equipartition analysis for TDEs where multi-frequency data is available and estimated launch dates of the radio outflows.  In \S\ref{sec:discussion} we discuss our findings in the context of the TDE population, with additional analysis presented in our companion paper, Alexander et al.~2023.  We summarize our conclusions in \S\ref{sec:conclusions}.

\section{Sample Selection \& Observations}
\label{sec:obs}

We observed 24 optically-selected TDEs, extending to $z \approx 0.16$, discovered between January 2014 and October 2020. The targets were drawn from the samples of \citet{vanvelzen2020} and \citet{Hammerstein2022}, supplemented with a few additional events from the literature. The sample of events is presented in Table~\ref{tab:properties}.  All TDEs were $2-6$ years post-discovery at the time of our observations.  We emphasize that one of these TDEs, AT2018hyz, showed an especially dramatic radio brightening starting at $\approx 3$ years post optical discovery and was presented in detail in a previous paper \citep{Cendes2022}.  The properties of all 24 TDEs in our sample are provided in Table~\ref{tab:obs}.

We obtained radio observations with the Karl G.~Jansky Very Large Array (VLA), the MeerKAT radio telescope, and the Australian Telescope Compact Array (ATCA).  VLA observations were first obtained in C-band (6 GHz), followed by multi-frequency observation in L- to K-band ($1-26$ GHz) in the event of a detection (Program IDs: 20A-492 PI: Alexander; 21A-303 PI: Hajela; 21B-360 PI: Cendes; 22B-205 PI: Cendes).  We processed the data using standard data reduction procedures in the Common Astronomy Software Application package (CASA; \citealt{McMullin2007}), using \texttt{tclean} on the calibrated measurement sets available in the NRAO archive.  We obtained all flux densities and uncertainties using the {\tt imtool fitsrc} command within the {\tt pwkit} package \citep{Williams2017}\footnote{https://github.com/pkgw/pwkit}. We assumed a point source fit, as preferred by the data. The resulting flux density measurements are provided in Table~\ref{tab:obs}. 

We obtained MeerKAT observations in L-band (1.36 GHz) and U-band (0.88 GHz; Program IDs: DDT-20200911-YC-01 PI: Cendes; SCI-20210212-YC-01 PI: Cendes; DDT-20220414-YC-01 PI: Cendes; SCI-20220822-YC-01 PI: Cendes; SCI-20220822-MB-03 PI: Bietenholz).  We used the standard calibrated MeerKAT pipeline images available via the SARAO Science Data Processor (SDP)\footnote{https://skaafrica.atlassian.net/wiki/spaces/ESDKB/pages/338723406/}, with the exception of data from SCI-20220822-MB-03 which was processed by {\tt OxKAT} \citep{Heywood2020}.  We confirmed via the secondary SDP products that the other sources in the MeerKAT images also in the NRAO VLA Sky Survey \citep[NVSS; ][]{Condon1998} were within $\sim10\%$ in expected flux.  We then obtained all flux densities and uncertainties using the {\tt imtool fitsrc} command within {\tt pwkit}. The resulting flux density measurements are provided in Table~\ref{tab:obs}. 

We obtained ATCA observations in S to Ku band ($2-20$ GHz; Program IDs: C3472 PI: Cendes; C3325 PI: Alexander).  We analyzed the data using the {\tt MIRIAD} package with the respective calibrators for absolute flux-density and band-pass, and to correct short term gain and phase changes. The {\tt invert}, {\tt mfclean} and {\tt restor} tasks were used to make deconvolved wideband, natural weighted images in each frequency band. Fluxe densities were determined by fitting a point source model using the {\tt Miriad imfit} tasks; for any upper limits, the image value at the expected location was used. The resulting flux density measurements are provided in Table~\ref{tab:obs}. 

Additionally, we obtained archival images from the Very Large Array Sky Survey \citep[VLASS; ][]{Lacy2020}, and from the Variables and Slow Transients Survey \citep[VAST; ][]{Murphy2021} where available; VLASS images are taken in S-band (3 GHz) and VAST images in UHF band (0.88 GHz).  We measured the flux densities in these images with the {\tt imtool fitsrc} command within {\tt pwkit}.  We also checked the NRAO data archive, MeerKAT data archive, and the Australia Telescope National Facility (ATNF) Data Archives and include any unpublished observations of our TDE sample, with the relevant project codes listed in Table~\ref{tab:obs}.

For all TDEs, we define the times of our observations relative to the date of the TDE discovery ($\delta t$). In the case of a non-detection, we report a $3\sigma$ upper limit.  We note that the uncertainties listed in Table~\ref{tab:obs} are statistical only and do not include an expected $\approx 3-5\%$ systematic uncertainty in the overall flux density calibration; we account for this systematic uncertainty in our subsequent modeling (\S\ref{sec:sed}.)

%\begin{landscape}
\startlongtable
\begin{deluxetable}{lcrlll}
\tablecolumns{6}
\tablecaption{TDEs Studied in this Work}
\tablehead{
Object&
$z$ &
Distance &
Discovery Date &
TDE Class &  
Discovery \\
  &
  &
(Mpc) &
 (UT) &
&  
Paper \\
}
\startdata
\multicolumn{6}{c}{\emph{TDEs with Previously-Known Radio Emission}} \\   
iPTF16fnl & 0.0163 & 71 & 2016 Aug 29 & TDE-H+He & \citet{Blagorodnova2017}\\
AT2019dsg & 0.0512 & 230 & 2019 Apr 9 &TDE-H+He & \citet{Hammerstein2023}\\
AT2020mot$^*$ & 0.070 & 317 & 2020 Jun 14 & TDE-H+He &\citet{Hammerstein2023} \\
\hline
\multicolumn{6}{c}{\emph{TDEs with Newly-Detected Radio Emission}} \\  
ASASSN-14ae & 0.0436 & 200 & 2014 Jan 25 &TDE-H+He$^{\dagger}$ & \citet{Holoien2014}\\
PS16dtm & 0.0804 & 368 & 2016 Aug 12 & TDE-H & \citet{Blanchard2017}\\
OGLE17aaj$^*$ & 0.116 & 540 & 2017 Jan 2 & TDE-H+He & \citet{Gromadzki2019}\\
AT2018zr & 0.071 & 323 & 2018 Mar 2 & TDE-H+He$^{\dagger}$ & \citet{vanVelzen2021}\\
AT2018dyb & 0.018 & 79 & 2018 Jul 11 & TDE-Bowen & \citet{Holoien2020}\\
AT2018bsi$^*$ & 0.051 & 228 & 2018 Apr 9 & TDE-H+He & \citet{vanVelzen2021}\\
AT2018hco & 0.088 & 404 & 2018 Oct 4 & TDE-H  & \citet{vanVelzen2021}\\
AT2018hyz & 0.046 & 205 & 2018 Oct 14 & TDE-H+He$^{\dagger}$  & \citet{Gomez2020,Short2020}\\
AT2019ehz & 0.074 & 338 & 2019 Apr 29 & TDE-H & \citet{vanVelzen2021}\\
AT2019eve & 0.081 & 372 & 2019 May 5 &TDE-H & \citet{vanVelzen2021}\\
AT2019teq & 0.087& 404 & 2019 Oct 20 &TDE-H+He &  \citet{Hammerstein2023}\\
AT2020nov$^*$ & 0.084&  385 & 2020 Jun 27 & TDE-H+He  & \citet{Frederick2020}\\
AT2020neh & 0.062 & 280 & 2020 Jun 19 & TDE-H+He & \citet{Angus2022}\\
AT2020pj$^*$ & 0.068 & 308 & 2020 Jan 2 &TDE-H+He & \citet{Hammerstein2023}\\
AT2020wey$^*$ & 0.027& 120 & 2020 Oct 8 & TDE-H+He &\citet{Hammerstein2023} \\
\hline
\multicolumn{6}{c}{\emph{TDEs with No Detected Radio Emission}} \\  
DES14C1kia & 0.162& 782 & 2014 Nov 11 & TDE-He & \citet{Foley2015}\\ %Auchettl2017
iPTF15af & 0.079 & 360 & 2015 Jan 15 & TDE-H+He & \citet{Blagorodnova2019}\\%,VanVelzen2020}\\
iPTF16axa & 0.108 & 503 & 2016 May 29 & TDE-H+He &  \citet{Hung2017}\\
AT2017eqx & 0.109 & 508 & 2017 Jun 7 & TDE-H+He &  \citet{Nicholl2019}\\
AT2018fyk & 0.059 & 264 & 2018 Sept 8 & TDE-H+He & \citet{Wevers2019}\\
AT2018lna & 0.091 & 419 & 2018 Dec 28 & TDE-H+He & \citet{vanVelzen2021}\\
\enddata
\tablecomments{For this table and throughout this paper we assume a flat $\Lambda$CDM cosmology with H$_0$ = 69.6 km s$^{-1}$ Mpc$^{-1}$, $\Omega_m$= 0.286, and $\Omega_\Lambda$=0.714 \citep{Wright2006}.\\
$^{*}$ TDEs for which radio emission was detected in our observations, but the nature of the emission is ambiguous, or due to star formation or AGN (see \S\ref{sec:ambiguous}).\\
$^{\dagger}$These TDEs showed only H lines in their initial classification spectra, leading to their labeling as TDE-H in some references. However, \citet{Holoien2014} report the emergence of He II in ASASSN-14ae in later epochs, which eventually becomes stronger than the H Balmer lines. Similarly, although AT2018hyz was classified by \citet{vanVelzen2021} as a TDE-H, \citet{Short2020} report the presence of both He I and He II emission lines (the latter appearing at $\sim70-100$ d). AT2018zr had the most delayed appearance of He II emission, at $\approx 170$ d post-discovery \citep{Hung2019}. We therefore list all three objects as TDE-H+He.}
\label{tab:properties}
\end{deluxetable}
%\end{landscape}
\section{Results}
\label{sec:results}

\subsection{Radio Luminosity and Time Evolution}

Of the 24 TDEs in our full sample (including AT2018hyz), 17 are detected in our observations on a timescale of $\approx 770-3250$ days. A detailed description of each TDE is provided in \S\ref{sec:descriptions}.  Of the 17 detected TDEs, 3 were detected in the radio at earlier times: iPTF16fnl \citep{Horesh2021b}, AT2019dsg \citep{Cendes2021b}, and AT2020mot \citep{Liodakis2022}.  Additionally, in 6 of the 17 detected TDEs the nature of the radio emission is ambiguous, namely they do not exhibit significant time evolution and/or lack constraining early limits; the radio emission in these cases may be due to a pre-existing AGN or star formation in the host galaxy.  This leaves 10 TDEs with definitive late-time radio emission well in excess of non-detections at early times (including AT2018hyz), requiring significant brightening hundreds of days post optical discovery. 

\begin{figure}
\centering
    \includegraphics[width=.85\columnwidth]{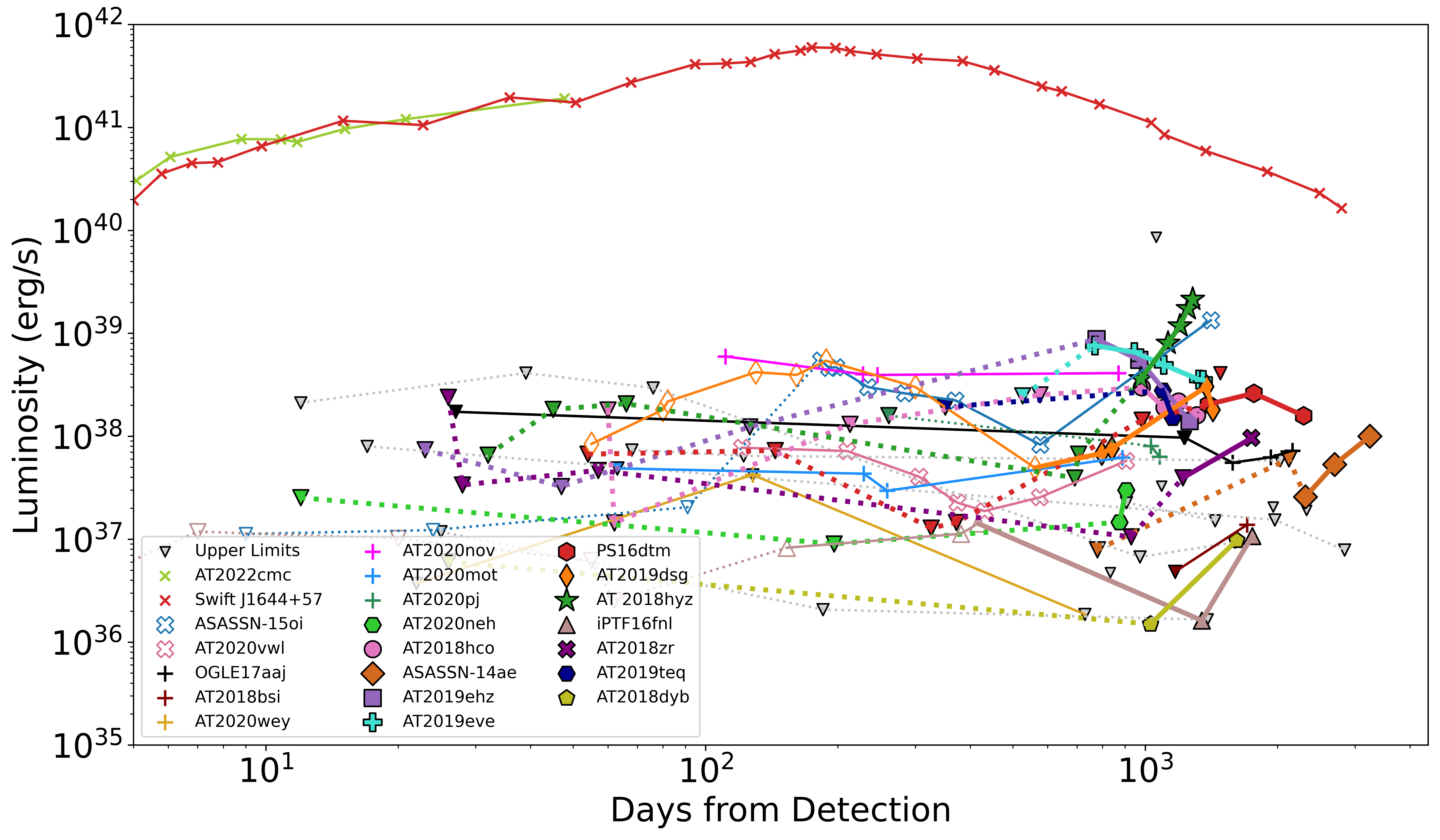}
        \includegraphics[width=.85\columnwidth]{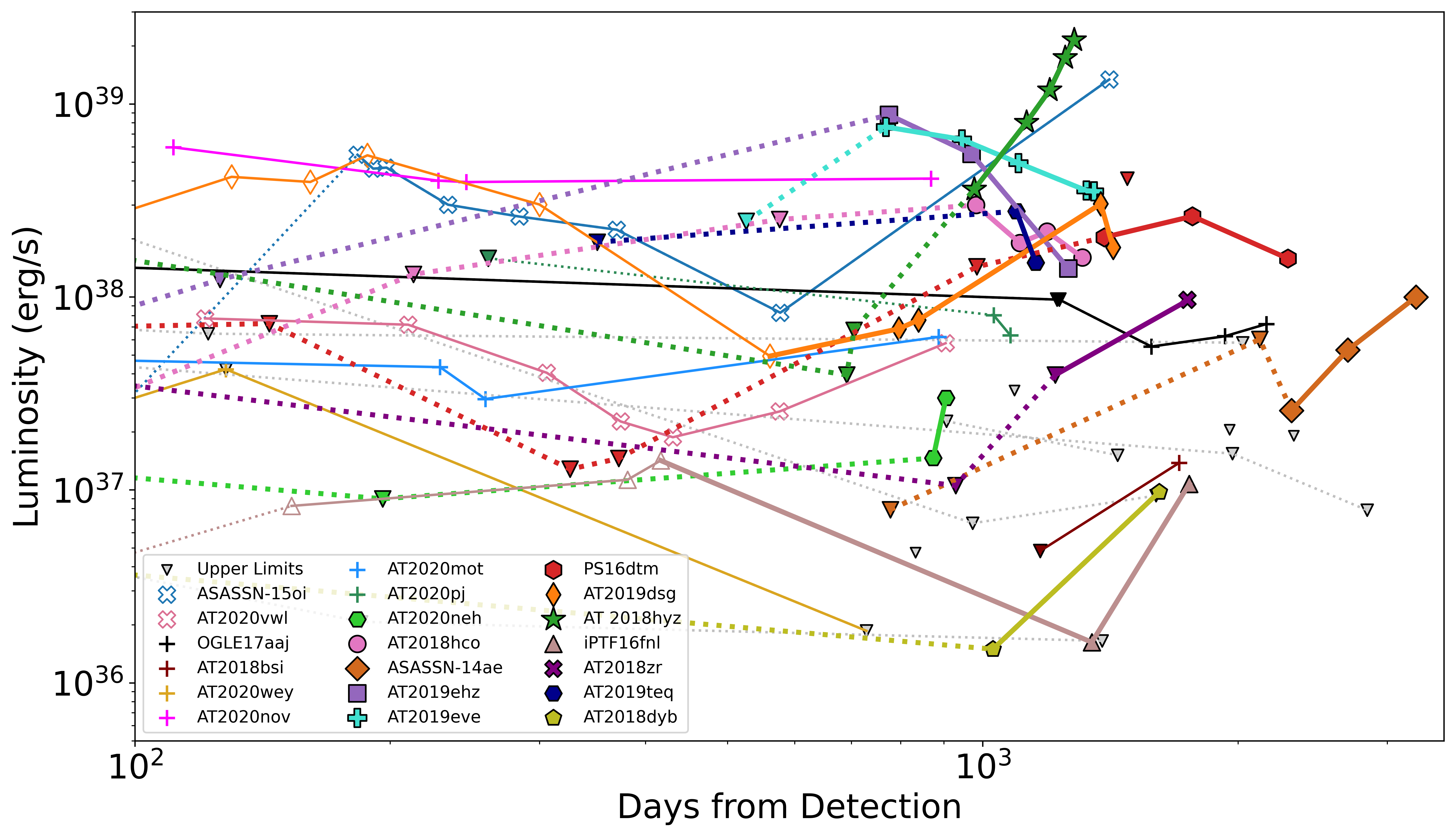}
    \label{fig:lumin-tde}
    \caption{\textit{Top:} Radio luminosity light curves for TDEs presented in this work (triangles: $3\sigma$ upper limits;  other symbols: detections).  All observations for the same TDE are connected with a dotted line for non-detections, and a solid line when detected.  TDEs with detected radio emission whose origin is ambiguous are shown as plus symbols (see \S\ref{sec:ambiguous}). We also include the light curve for AT2018hyz from \citet{Cendes2022}. For comparison we also show radio light curves for TDEs with early jetted radio emission (Sw1644+57: \citealt{Cendes2021}; AT2022cmc: \citealt{Andreoni2022}) and TDEs with late brightening (ASASSN-15oi: \citealt{Horesh2021}; AT2020vwl: \citealt{Goodwin2023,Goodwin2023C}) as well as one TDE with early radio emission for which we detect significant re-brightening (\citealt{Cendes2021b,Stein2021}), where  previously published data are shown as open symbols, and our new data with filled symbols connected by thicker lines.  We do not plot non-constraining upper limits, but they are available in Table~\ref{tab:obs}.  \textit{Bottom:} the same data presented above, but zoomed in to only show observations at $>100$ d, and luminosities of $< 3\times10^{39}$ erg s$^{-1}$, highlighting the significant population of TDEs with late-rising radio emission.}
\end{figure}

We present the radio light curves for all 24 sources in this full sample, including upper limits, in Figure~\ref{fig:lumin-tde}. For the purpose of comparison, we also include the radio light curves of the jetted TDEs SwJ1644+57 \citep{Eftekhari2018,Cendes2021} and AT2022cmc \citep{Andreoni2022}, and two previous TDEs with late radio emission, ASASSN-15oi \citep{Horesh2021} and AT2020vwl \citep{Goodwin2023,Goodwin2023C}. We also include previously published data for iPTF16fnl \citep{Blagorodnova2017,Horesh2021b}, AT2018dyb \citep{Holoien2020}, iPTF15af \citep{Blagorodnova2019}, PS16dtm \citep{Blanchard2017}, AT2018hco \citep{Horesh2018}, AT2020mot \citep{Liodakis2022}, AT2018hyz \citep{Cendes2022} %AT2022cmc \citep{Andreoni2022}, 
and AT2019dsg \citep{Stein2021,Cendes2021b}. In all cases where multi-frequency observations are available, we use data in C-band (6 GHz).

\subsubsection{TDEs with Newly-Discovered Late Radio Emission}
\label{sec:descriptions}

\begin{figure}
\centering
        \includegraphics[width=.85\columnwidth]{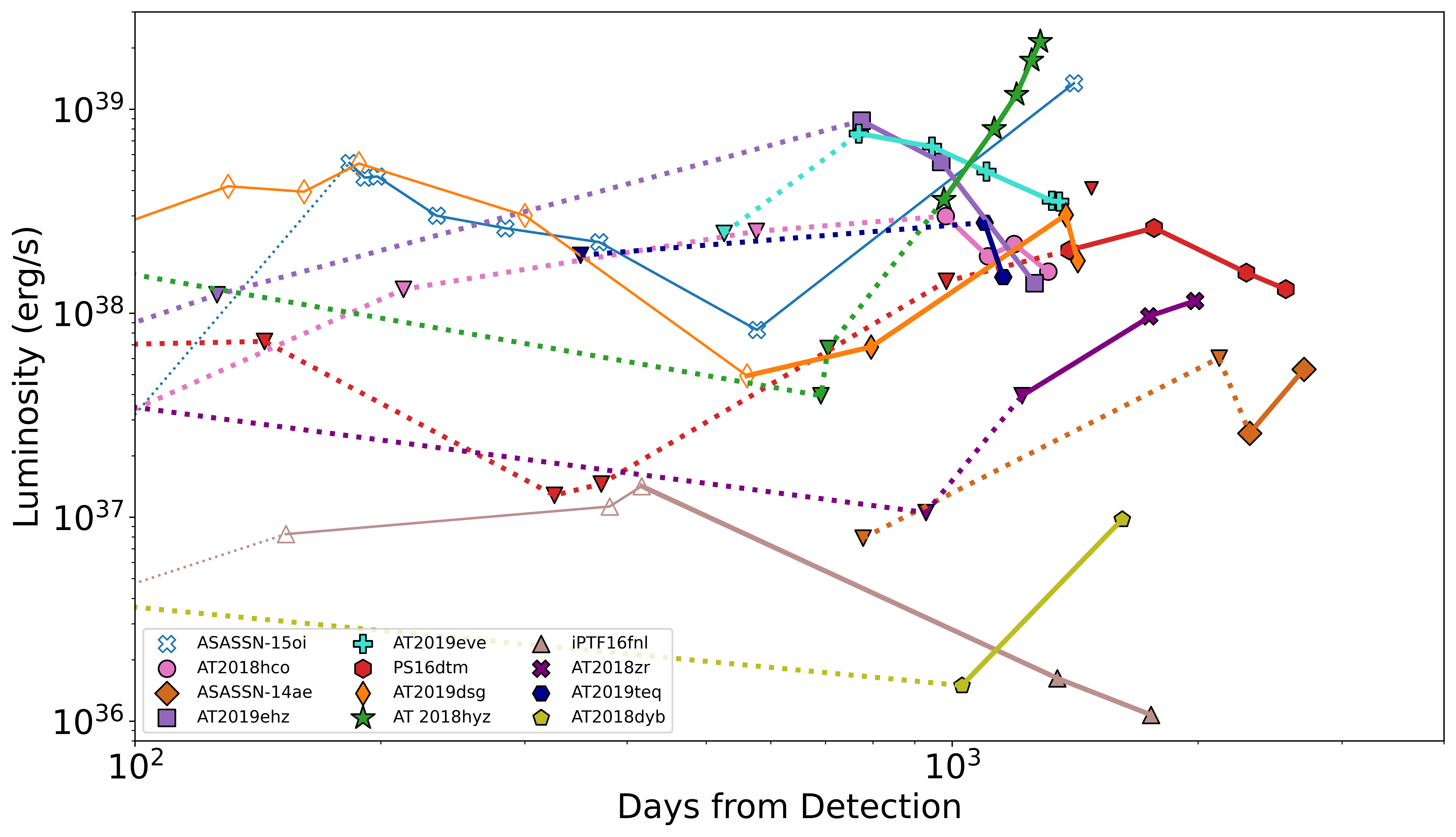}
    \label{fig:lumin-subset}
    \caption{As in the bottom of Figure \ref{fig:lumin-tde}, but only for the main sources listed in Sections \ref{sec:descriptions} and \ref{sec:previous}}.
\end{figure}

We report 9 new TDEs with late-time radio emission identified based on our observations and constraining earlier non-detections (from targeted or survey data).  These light curves can be seen in Figure \ref{fig:lumin-subset}.  Below we briefly describe the radio light curve properties of each event.

\begin{itemize}
    \item \textbf{ASASSN-14ae} was optically discovered on 2014 January 25 at a redshift of $z=0.0436$ \citep{Holoien2014}, making it the oldest TDE in our sample.  A VLA observation at 778 days yielded an upper limit of $\lesssim 0.033$ mJy at 6 GHz, and a VLASS observation at 2122 days yielded an upper limit of $\lesssim 0.42$ mJy at 3 GHz.  We first detected ASASSN-14ae at 2313 days when it was $0.090\pm 0.015$ mJy at 6 GHz, and found a steady (and on-going) rise in its light curve over subsequent observations to $0.42\pm 0.08$ mJy at 3243 days, or a luminosity of $1\times10^{38}$ erg s$^{-1}$. This corresponds with a factor of $\sim4$ increase in luminosity since detection, with a steep power law rise since initial detection of ($F_\nu\propto t^\alpha$) with $\alpha= 4.2$. Continued observations of the rising light curve are on-going.
    
    \item \textbf{PS16dtm} was optically discovered on 2016 August 12 at a redshift of $z=0.0804$ \citep{Blanchard2017,Jiang2017,Petrushevska2023}.  Four radio observations in the first $\sim{\rm year}$ resulted in non-detections, with the first non-detection at 54 days and the latest non-detection at 372 days, when it was $\lesssim 0.015$ mJy at 6 GHz \citep{Blanchard2017}.  It should be noted that PS16dtm had a pre-existing AGN; however with several non-detections in radio it is expected that the AGN contribution to radio emission was minimal \citep{Blanchard2017}.  We first detected PS16dtm at 1391 days with $0.210\pm0.008$ mJy at 6 GHz ($2.0\times10^{38}$ erg s$^{-1}$).  The light curve rose to a peak of $0.271\pm0.002$ mJy ($2.6\times10^{38}$ erg s$^{-1}$) at 1767 days, and faded to $0.163\pm0.012$ mJy ($1.6\times10^{38}$ erg s$^{-1}$) by 2291 days.
    
    \item \textbf{AT2018zr} (also known as PS18kh) was discovered optically on 2018 March 2 at $z=0.075$ \citep{Hung2019}. Initial observations with the Arcminute Microkelvin Imager (AMI) at 16 GHz and the VLA at 10 GHz at $26-57$ days yielded no detections \citep{vanVelzen2019}; observations from our program at 929 and 1218 days at 6 GHz yielded upper limits of $<0.014$ and $<0.053$ mJy, respectively.  The source was then detected at 1713 days, with $0.147\pm 0.011$ mJy at 6 GHz, and had risen to $0.155\pm 0.018$ mJy 30 days later.  This corresponds to a rise in luminosity of $\gtrsim2.5\times$ from the last upper limit ($\lesssim 3.9\times10^{37}$ erg s$^{-1}$ to $9.7\times10^{37}$ erg s$^{-1}$), corresponding to a power law index of $\alpha\gtrsim2.7$ from its last upper limit to first detection.  The TDE was at peak luminosity in our last observation, and future observations will allow us to determine the evolution of this TDE.
    
    \item \textbf{AT2019teq} was optically discovered on 2019 October 20 at $z= 0.087$ \citep{Hammerstein2022}.  \citet{Yao2022} reported an X-ray brightening and hardening of this TDE on 2022 September 8 (1050 days), which was confirmed by NICER observations on 2022 October 18-21 (1100 days).  No prior radio observations of AT2019teq exist except for a VLASS observation at 351 days with an upper limit of $<0.33$ mJy at 3 GHz.  Our VLA observation at 1096 days at 6 GHz resulted in a detection with $0.238\pm 0.008$ mJy \citep{ATelCendes2022}, and the emission subsequently faded at 1155 days to $0.129\pm 0.016$ mJy.  Given the decline between our two observations, and the earlier non-detection, we conclude that the radio emission peaked at $\sim 400-1000$ days; the lower limit on the peak radio luminosity is $\approx 2.7\times 10^{38}$ erg s$^{-1}$.
    
    \item \textbf{AT2018dyb} (ASASSN-18pg) was optically discovered on 2018 July 11, at $z = 0.018$ \citep{Holoien2020,Leloudas2019}.  Radio observations at 26 days led to an upper limit of $<0.43$ mJy at 19 GHz.  Our first observation took place at 1028 days and led to a detection with $0.158\pm 0.06$ mJy at 1.3 GHz.  The emission then dramatically brightened at 1615 days to $1.03\pm 0.07$ mJy ($\approx 10^{37}$ erg s$^{-1}$), corresponding to a power law index of $\alpha\approx4.2$ since its first detection. Future observations will allow us to determine the evolution of this TDE.
    
    \item \textbf{AT2018hco} was optically discovered on 2018 October 4 at $z=0.088$ \citep{vanvelzen2020}.  Radio non-detections were obtained with AMI at 15.5 GHz at 60 days \citep[$<0.08$ mJy; ][]{Horesh2018}.  We also identified a VLA archival observation at 62 days leading to a limit of $<0.016$ mJy at 6 GHz (18A-373 PI: van Velzen), a VLASS observation at 213 days with $<0.30$ mJy, and a limit of $<1.986$ mJy at 0.88 GHz with the ASKAP VAST survey.  We first detected radio emission at 982 days with $0.343\pm0.015$ mJy at 6 GHz, or a luminosity of $4\times10^{38}$ erg s$^{-1}$. The emission then fades in subsequent observations to $0.265\pm0.014$ mJy at 5 GHz at 1191 days, and we then see the source fading over time, to $0.200\pm0.019$ mJy at 5.5 GHz with ATCA on 1311 days ($1.6\times10^{38}$ erg s$^{-1}$).  We note the dip in luminosity at 1106 days is due to the use of VLASS observation at 3 GHz compared to 6 GHz for the other data points.  The decline in luminosity from the first to the last detection has a power law index of $\alpha\approx -1.8$.  We estimate that the light curve peak occurred at $\approx220-980$d.  
    
    \item \textbf{AT2019ehz} was optically discovered on 2019 April 29 at $z = 0.074$ \citep{vanVelzen2021}.  We obtained early limits from unpublished archival VLA data at 23 and 47 days (19A-395; PI: van Velzen) with resulting limits of $<0.06$ and $<0.26$ mJy, respectively, at 9 GHz.  There is also a VLASS non-detection at 126 days with $<0.30$ mJy.  We first detected radio emission at 775 days, with $1.07\pm 0.003$ mJy at 6 GHz, or a luminosity of $8.7\times10^{38}$ erg s$^{-1}$. The emission subsequently faded in observations at 970 and 1262 days, to $0.205\pm 0.034$ mJy, corresponding to a power law decline of $\alpha\approx -3.4$.  Since the emission is declining and the last upper limit before detection is at 126 days, we conclude that the light peak occurred at $\sim 130-700$ days. However, we note with the inferred steep decline in luminosity, it is more likely the peak was actually much closer to the time of first detection.  This could have implications for the launch of the outflow; for example, if the time of the delayed launch was at $\approx 600$ days the power law index would be about -1.  We discuss this further in Section~\ref{sec:outflow-date}.
    
    \item \textbf{AT2019eve} was optically discovered on 2019 May 5 at $z=0.081$ \citep{vanVelzen2021}. A VLASS observation at 526 days yielded a non-detection of $<0.50$ mJy.  We detected this source at 769 days with $0.766\pm0.009$ mJy at 6 GHz, or a luminosity of $7.6\times10^{38}$ erg s$^{-1}$.  Subsequent observations to $1353$ days indicate steady fading to a final flux density of $0.711\pm 0.133$ mJy.  The power law index for the rise between the last upper limit and first detection is $\alpha\gtrsim 3$, while for the decline it is $\alpha\approx -1.4$.  The light curve peak occurred at $\approx 530-770$ days.
    
    \item \textbf{AT2020neh} was optically detected on 2020 June 19 at $z=0.062$ \citep{Angus2022}.  Radio observations at 12 and 196 days led to non-detections with $<0.018$ mJy and $<0.016$ mJy, respectively, at 15 GHz \citep{Angus2022}. Our first detection is at 874 days with $0.026\pm 0.006$ mJy at 6 GHz, followed about 30 days later by a rise to $0.053\pm 0.012$ ($\approx 3\times 10^{37}$ erg s$^{-1}$), or $\alpha\approx 20$.  Despite the apparent rapid rise, the faintness of the emission precluded multi-frequency observations. Future observations will allow us to determine the evolution of this TDE.
\end{itemize}

\subsubsection{TDEs with Previously-Known Radio Emission}
\label{sec:previous}

We identify a distinct late-time radio re-brightening in one TDE that exhibited early radio emission with fading behavior prior to our observations, and one TDE that was identified in the literature as a late-time radio TDE \citep{Horesh2021b}.  We also include these light curves in Figure \ref{fig:lumin-subset}.  Below we briefly describe the radio light curve properties of each event.

\begin{itemize}
    \item \textbf{iPTF16fnl} was optically discovered on 2016 August 29 at $z = 0.0163$ \citep{Blagorodnova2017}.  There were several radio non-detections at $2-62$ days with $\lesssim 0.027-0.12$ mJy at 15 GHz ($\lesssim 2.5\times10^{36}-1.2\times10^{37}$ erg s$^{-1}$), followed by detections at 15 GHz at $153-417$ days, with a peak luminosity at 417 days of $\approx 1.4\times10^{37}$ erg s$^{-1}$ \citep{Horesh2021b}.  We observed iPTF16fnl at 1345 days at 6 GHz and found that the source had faded to $0.045\pm 0.001$ mJy, $\approx 1.6\times10^{36}$ erg s$^{-1}$. Our subsequent observation at 1752 days shows the source has faded to $0.0295\pm0.007$ mJy at 6 GHz ($1.1\times10^{36}$ erg s$^{-1}$).
    
    \item \textbf{AT2019dsg} was optically discovered on 2019 April 9 at $z = 0.051$ \citep{vanvelzen2020}. Radio emission was first detected at 52 days, steadily rose to a peak at $\approx 200$ days with a luminosity of $\approx 5.4\times10^{38}$ erg s$^{-1}$, and then steadily declined through 560 days, to $\approx 4.9\times10^{37}$ erg s$^{-1}$ \citep{Cendes2021b,Stein2021}.  Our new VLA observations at 796 days revealed a slight re-brightening, with $\approx 6.8\times10^{37}$ erg s$^{-1}$ at 6 GHz; this is about an order of magnitude brighter than expected from continued steady decline. A follow-up MeerKAT observation at 1378 days had a flux density of $0.384\pm0.026$ mJy at 1.36 GHz.  Extrapolating to 6 GHz assuming the same spectral energy distribution as observed  at 1170 days leads to an estimated flux density of $\approx 0.8$ mJy ($\approx 3\times10^{38}$ erg s$^{-1}$), or a power law index of $\alpha\approx 3$ during this time period.  A final observation on 1437 days at 1.36 GHz indicated the source had faded to $0.180\pm0.009$ mJy, extrapolated to 6 GHz as $\approx 0.48$ mJy ($\approx 2\times10^{38}$ erg s$^{-1}$).  Thus, we conclude that AT2019dsg has evidence for a separate emission component than its initial peak; continued observations will delineate this TDE's time evolution.
\end{itemize}

\subsubsection{TDEs with Ambiguous or Host/AGN Radio Emission}
\label{sec:ambiguous}

We identify radio emission in an additional 6 TDEs, but we cannot definitively ascertain its nature, due to an absence of earlier deep upper limits or a lack of significant time evolution during our observations.  We exclude these sources from subsequent detailed analysis of our TDE sample and treat them in \S\ref{sec:rate} as upper limits. Below we provide relevant information for completeness; the radio data for these TDEs are provided in Table~ \ref{tab:obs-not-used}.

\begin{itemize}
    \item \textbf{OGLE17aaj} was discovered on 2017 January 2 at $z = 0.116$ \citep{Gromadzki2019}.  We detect this source for the first time with MeerKAT at 1581 days at 1.6 GHz, with a flux density of $0.19\pm0.02$ mJy. We have also identified observations of this source as part of the ASKAP VAST survey at 1.42 GHz at 1224 and 1231 days, but these lead to non-constraining upper limits of $\lesssim 0.3$ mJy ($3\sigma$).  In subsequent observations with MeerKAT and ATCA at 2161 days we find that the radio emission remains fairly steady ($\alpha\propto 0.6$).  This, combined with the lack of a constraining upper limit at earlier times, leads us to conclude that the radio emission from OGLE17aaj is unlikely to be related to the TDE.
    
    \item \textbf{AT2018bsi} was optically discovered on 2018 April 9 at $z = 0.051$ \citep{vanvelzen2020}.  We first observed this source with the VLA at 1169 days at 6 GHz, and found an upper limit of $\lesssim 0.013$ mJy.  A subsequent observation at 1705 days led to a significant detection at 6 GHz, with $0.037\pm 0.005$ mJy.  We note that VLASS observations, on May 24, 2019 and October 26, 2021 respectively, were non-constraining upper limits of $<0.3$ mJy.   However, due to the faintness of the emission, we are presently unable to obtain multi-frequency follow-up, and we therefore classify the radio emission as ambiguous until additional observations can be taken.

    \item \textbf{AT2020pj} was optically discovered on 2020 January 2 at $z = 0.068$ \citep{Hammerstein2022}. The first available radio observation was at 261 d from VLASS, leading to an upper limit of $\lesssim 0.465$ mJy at 3 GHz. We subsequently detected the TDE at 6 GHz on 1030 d with a flux density of $0.118\pm 0.006$ mJy, and on 1078 d with $0.093\pm 0.011$ mJy.  Due to the initial shallow upper limit, and the marginal variability we cannot determine if the radio emission is due to the TDE.  Continued monitoring will help to ascertain the nature of this emission.
    
    \item \textbf{AT2020mot} was optically discovered on 2020 June 14 at $z = 0.070$ \citep{Liodakis2022}. It was first observed at 67 days at 15 GHz, with an upper limit of $<0.027$ mJy \citep{Liodakis2022}, but was subsequently detected at 229 days  with $0.060\pm0.008$ mJy and 259 days with $0.041\pm 0.008$ mJy at 6 GHz by \citet{Liodakis2022} who concluded that the emission was likely due to star formation activity in the host galaxy.  We detected the source at 887 days with $0.086\pm 0.008$ mJy at 6 GHz, and conclude that the variations in flux are more consistent with an AGN than star formation activity, but likely unrelated to the TDE.
    
    \item \textbf{AT2020nov} was optically discovered on 2020 June 27 at $z = 0.084$ \citep{Frederick2020}.  Radio emission was first detected at 111 days at 15 GHz with a flux density of 0.224$\pm$0.009 mJy (VLA program 20A-372, PI: Alexander), and the source had subsequent detections at 228 and 246 days at 6 GHz, when it was 0.376$\pm$0.010 mJy and 0.370$\pm$0.01 mJy, respectively.  In our own program, we observed AT2020nov at $t_d= 869$d at 6 GHz and found the source to be $0.39\pm0.02$ mJy, which is consistent with a steady luminosity at earlier times.  Due to a lack of variability, we conclude this emission is likely due to non-TDE related emission in the host galaxy, such as star formation.

    \item \textbf{AT2020wey} was optically discovered on 2020 October 8 at $z = 0.027$ \citep{Hammerstein2023}.  A radio observation 22 days post-discovery had an upper limit of $<0.014$ mJy at 15 GHz, followed by a detection at 128 days at 6 GHz of $0.0408\pm 0.0081$ mJy (VLA program 20A-372, PI: Alexander).  Our program observed this source at 6 GHz at $t_d= 729d$, and yielded a non-detection of $<0.018$ mJy.  Due to the weak detection at 128 days, and a lack of follow-up detections, we cannot conclude the radio emission is due to the TDE.  We will continue monitoring this TDE to ascertain the nature of this emission in the future.
\end{itemize}

\subsection{The Incidence Rate and Properties of Late-Time Radio Emission}
\label{sec:rate}

\begin{figure}
\centering
    \includegraphics[width=.45\columnwidth]{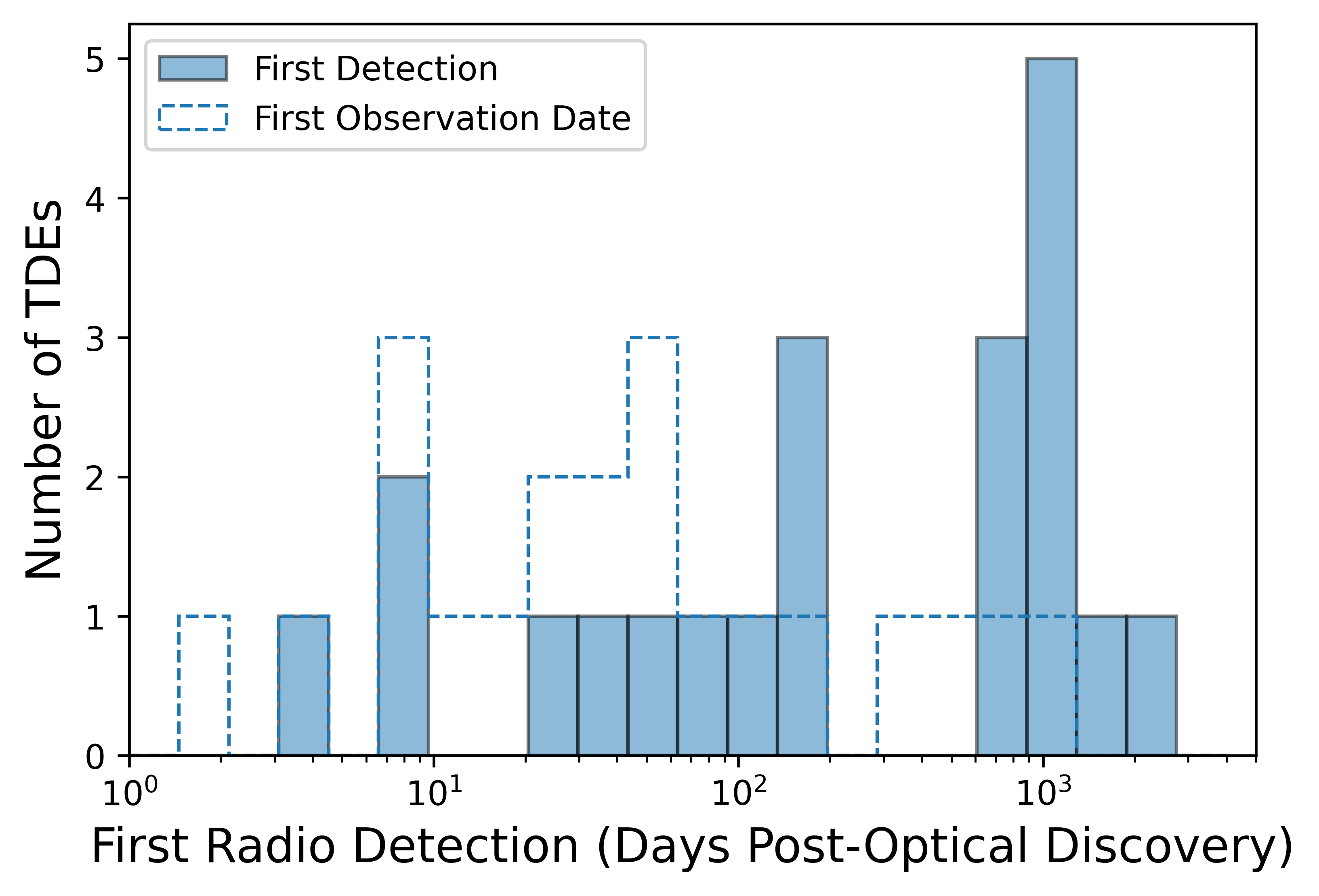}
    \includegraphics[width=.45\columnwidth]{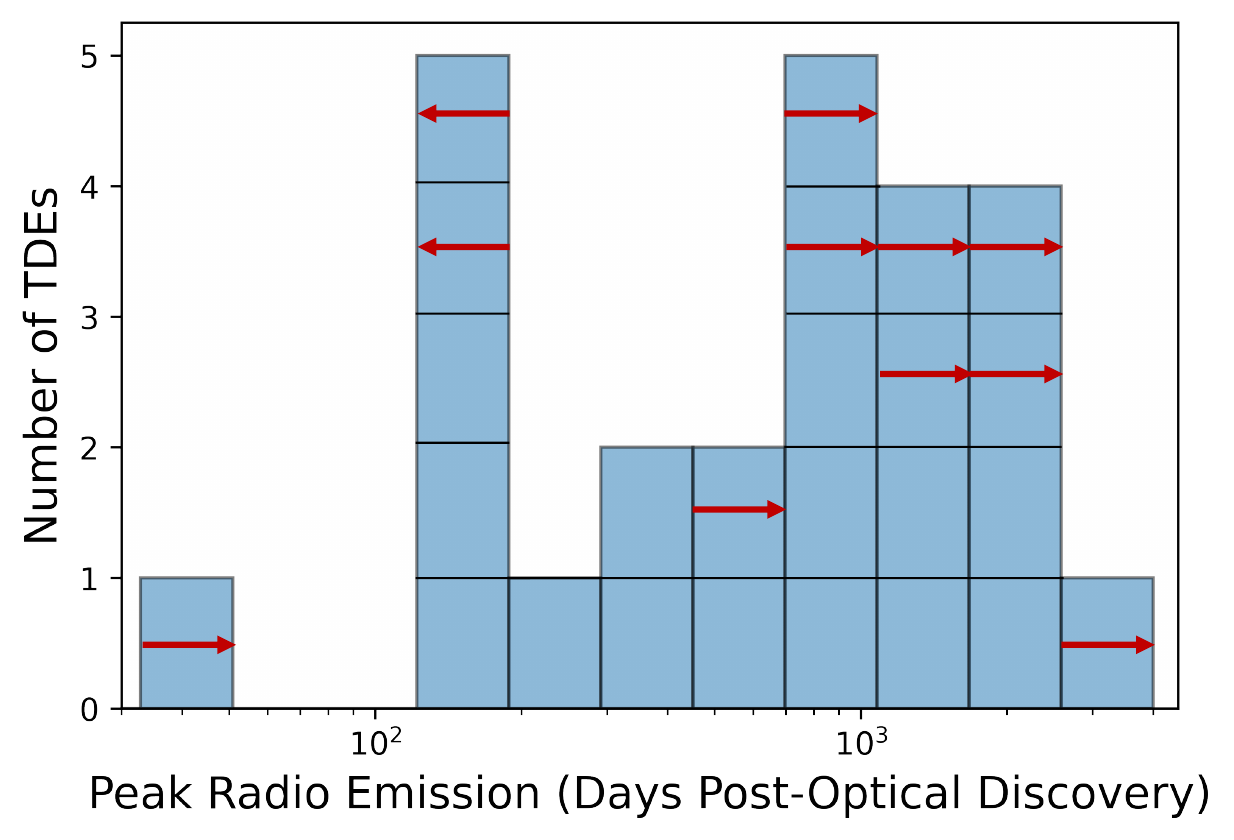}
    \includegraphics[width=.45\columnwidth]{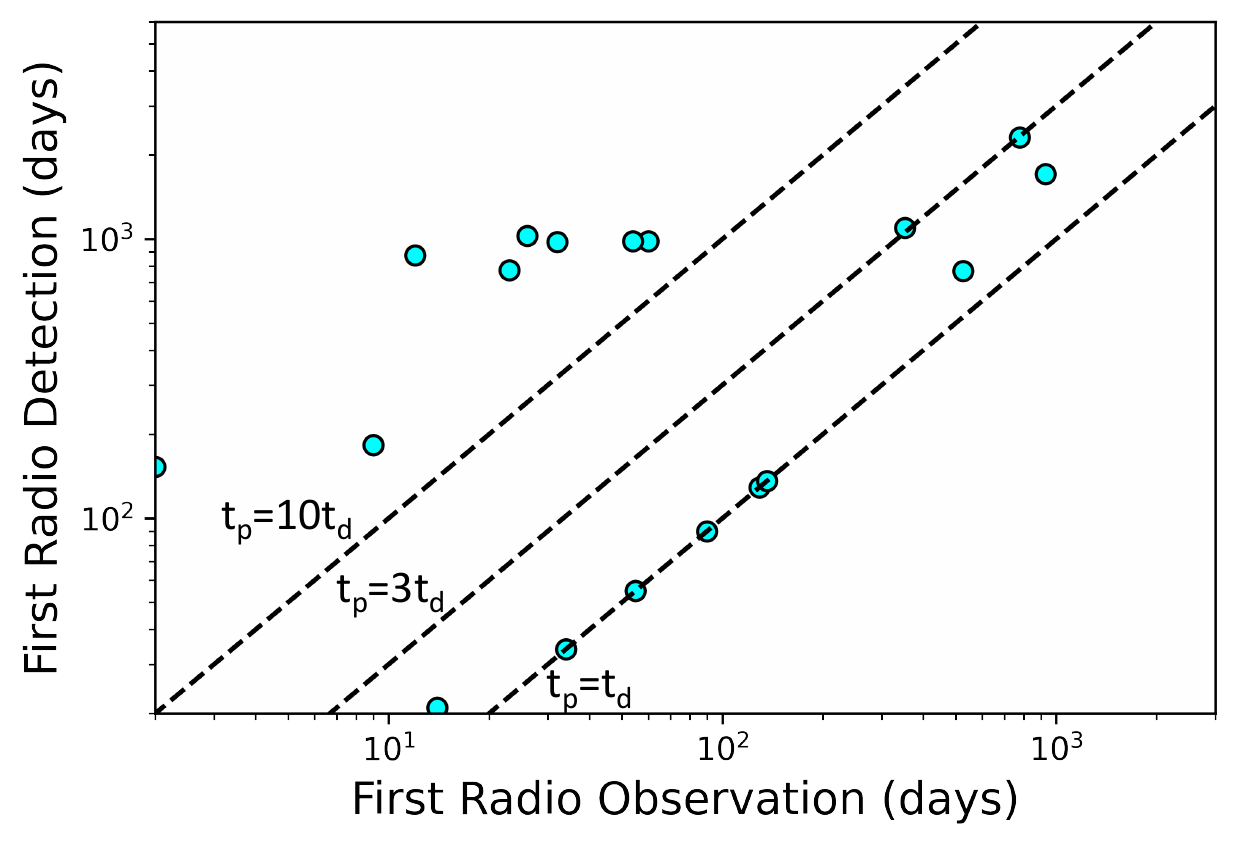}
    \includegraphics[width=.45\columnwidth]{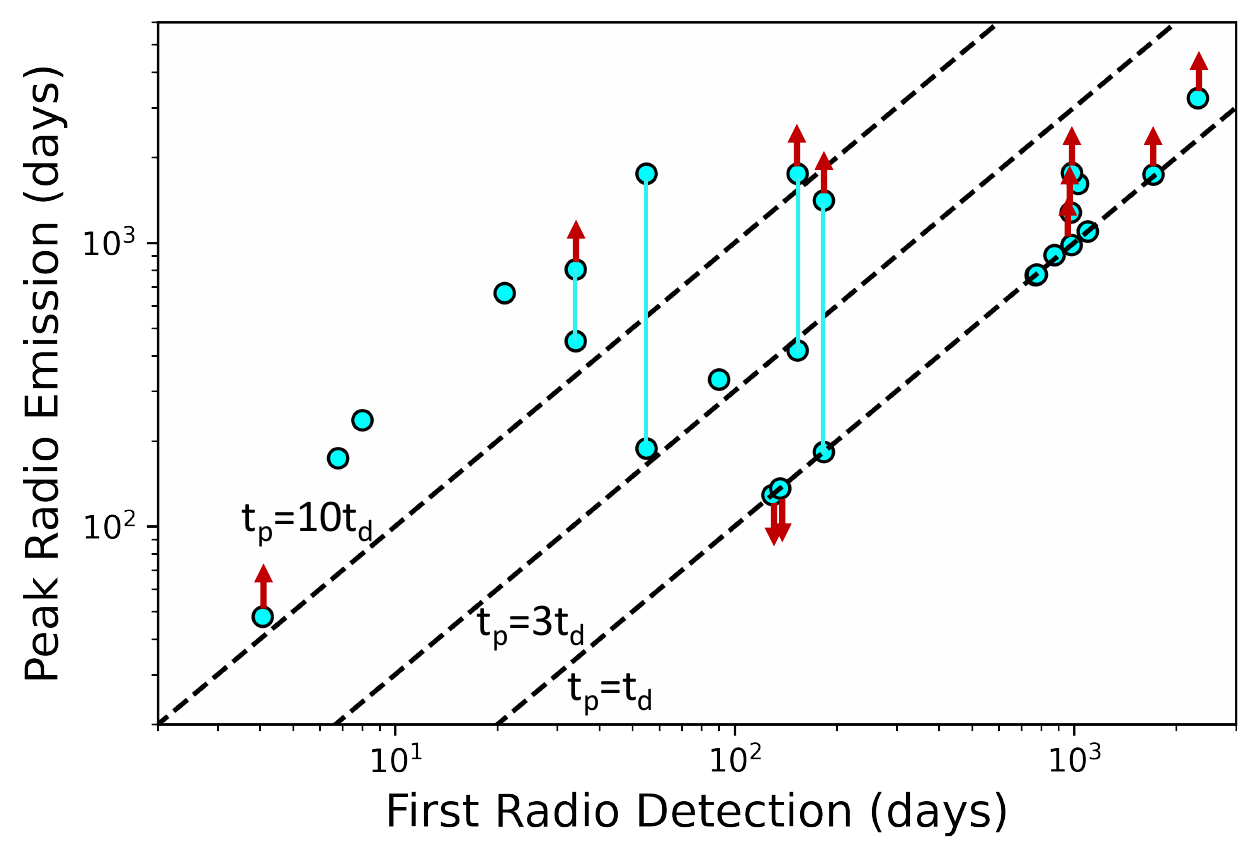}
    \label{fig:histo}
    \caption{\textit{Upper left:} Histograms of the time of first radio detection (solid) and first radio observation (dashed) for TDEs with detected radio emission. \textit{Upper right:} Histogram of peak radio emission timescale at $\sim 6$ GHz for TDEs with detected radio emission.  Arrows indicate upper and lower limits. For TDEs with multiple distinct peaks (ASASSN-15oi, AT2019dsg), we include both components. \textit{Bottom left:} The time of first radio observation versus time of first radio detection.  The diagonal lines mark first radio detection, $t_o$, at multiples of $1,3,10$ times the time of first observation.  \textit{Bottom right:} Time of peak radio emission versus time of first detection, with arrows indicating upper and lower limits; for the TDEs with distinct peaks (ASASSN-15oi, AT2019dsg, AT2020vwl) we include both components connected by a solid line.  The diagonal lines mark peak radio emission, $t_p$, at multiples of $1,3,10$ times the time of first detection. This indicates that for the TDE population with late radio emission at least some events may peak on a decade timescale.  In addition to the data presented in this paper, radio data are from: AT2019qiz \citep[][Alexander et al. in prep]{OBrien2019}, AT2019azh \citep{Goodwin2022,Sfaradi2022}, AT2019ahk (Christy et al.~in prep), AT2019dsg \citep{Cendes2021b,Stein2021}, AT2020opy \citep{Goodwin2023B}, ASASSN-14li \citep{Alexander2016}, AT2020vwl \citep{Goodwin2023,Goodwin2023C}, iPTF16fnl \citep{Horesh2021b}, and ASASSN-15oi \citep{Horesh2021}.  We also include the jetted TDEs AT2022cmc \citep{Andreoni2022}, Swift J1112.2-8238 \citep{Brown2017}, Swift J2058.4+0516 \citep{Cenko2012}, and Sw1644+57 \citep{Zauderer2011}.  We exclude TDEs in this plot where radio emission was ambiguous in nature (see \S\ref{sec:ambiguous}).}
\end{figure}

Our sample of TDEs with radio observations is the largest to date: we observed 24 optically-selected TDEs (of which AT2018hyz was the subject of a separate paper: \citealt{Cendes2022}).  In this paper we identify 9 new TDEs that had constraining radio upper limits at early times, and then exhibit brighter radio emission hundreds of days post optical discovery; this excludes the 6 events with radio emission that is either ambiguous in origin or unlikely to be associated with the TDE (\S\ref{sec:ambiguous}), and 2 events with prior radio emission (\S\ref{sec:previous}). Including AT2018hyz, this corresponds to a high detection fraction of $10/22$ or $\approx 45\%$.  Alternatively, if we count distinct late-time brightening in AT2019dsg we obtain a detection fraction of $11/24$ or $\approx 45\%$.  Thus, regardless of the exact accounting we conclude that about half of all optically-selected TDEs exhibit radio emission that rises on timescales of hundreds of days. This high fraction is particularly striking when compared to the published statistics of early radio detections of optically-selected TDEs ($\lesssim 200$ days) of $\approx 30\%$ \citep{Alexander2020}.

In Figure~\ref{fig:histo} we explore the turn-on and peak timescales of detected radio emission in the full TDE population with radio detections.  The left panel of Figure~\ref{fig:histo} shows the timescale at which radio emission is first detected. We find a broad range of timescales, spanning from a few days to $\approx 2300$ days. We note that some TDEs without current radio detections may yet turn on at even later timescales, as highlighted by the case of ASASSN-14ae with a first detection at $\approx 2300$ days, and is still rising.  The overall distribution of turn-on timescales appears to exhibit three groupings. First, at $\lesssim 20$ d are the jetted TDEs (Sw J1644+57, Swift J2058.4+0516, AT2022cmc), which are detected early due to the combination of rapid triggering and luminous radio emission from a relativistic jet, as well as the rapidly-evolving AT2019qiz, which was detected in the radio at 8 days \citep{Nicholl2020}.  Second, at $\approx 20-200$ days we find 8 TDEs\footnote{We note that this grouping includes ASASSN-15oi with a first detection at 183 days and iPTF16fnl with a first detection at 153 days, which were referred to as ``late'' emission by \citet{Horesh2021} and \citet{Horesh2021b}, respectively, but which we clearly see here are more typical of TDEs with early radio emission, and are distinct from the population of TDEs with radio emission only at $\gtrsim 10^3$ days identified here.}).  Finally, identified here for the first time, we find that about half of all TDEs with radio emission are detected only at $\gtrsim 600$ days, with a peak at $\sim 10^3$ days.  We note that the gap at $\approx 200-600$ days may be due to observing gaps, so it is possible that there is a more continuous distribution of turn-on times; however, it is clear that turn-on timescales of $\sim 10^3$ days are at least (or more) common than at $\sim 10^2$ days.  We also show in the left panel of Figure~\ref{fig:histo} the distribution of timescales of the first radio {\it observation} of each TDE. We note that for about $40\%$ of the TDEs, the first observation led to the first detection (e.g., ASASSN-14li, AT2019dsg; \citealt{Alexander2016,Stein2021}), and it is therefore likely that a first radio detection would have been possible even earlier; for the remaining TDEs there is at least one non-detection prior to the first detection.

While the turn-on timescale is informative, and clearly hints at a distinction between early- ($\sim 10^2$ d) and late-rising ($\sim 10^3$ d) radio emission, the time of first detection is at least in part affected by any delays in the first observation. An additional relevant timescale is that of peak radio emission, shown in the right panel of Figure~\ref{fig:histo}; we use the time of peak at $\sim 6$ GHz for uniformity.  We note that in some cases the radio light curve is already declining at the time of discovery so only an upper limit on the peak timescale is available; conversely, in other cases the emission is still rising in our latest observations leading to a lower limit on the peak timescale.  We are also further limited by the various detection limits and event distances in the TDE population.  We further note that for the 3 TDEs with a clear double-peaked structure (ASASSN-15oi, AT2020vwl and AT2019dsg) we include both peaks in the distribution. The overall distribution is somewhat more uniform than the distribution of first detections, but we still note a bimodality, with peaks at $\sim 150$ d and $\sim 1500$ d (especially when we consider upper and lower limits). Regardless of the exact structure of the distribution, we find that $\approx 50\%$ of TDEs with radio emission peak at $\gtrsim 10^3$ days.  This is further highlighted in the bottom panel of Figure~\ref{fig:histo}, where we plot the peak timescale versus the time of first detection, indicating that for events with early emission the peak emission timescale is typically $\sim 3-10$ times higher than the time of first detection; while the events with first detections at $\sim 10^3$ d have mostly not reached their peak, if they had similar ratios they would peak on $\gtrsim {\rm decade}$ timescales.

Investigating the radio luminosities of the late emission, we find a range of $\approx 10^{37}$ to $\gtrsim 2\times 10^{39}$ erg s$^{-1}$, but we stress that for several TDEs the emission is still rising, so both the lower and upper ranges may shift higher once all sources reach their peak.  Overall, these radio luminosities are comparable to those of TDEs with early radio emission, but are $\approx 30-3000$ times less luminous than Sw1644+57 at a comparable timescale. We note that in the cases where the peak luminosity is well constrained by our detections and preceding upper limits (i.e., PS16dtm, AT2019eve, AT2019ezh, AT2018hco) the peak luminosities are $\lesssim 10^{39}$ erg s$^{-1}$, compared to $\approx 10^{41}$ erg s$^{-1}$ for Sw1644+57, and it is therefore unlikely that their radio emission is due to initially off-axis jets; we return to this point in \S\ref{sec:jet}.  On the other hand, the radio emission is still rising in some TDEs (i.e., ASASSN-14ae, AT2018dyb, AT2018zr, AT2019dsg) although we note that those too currently have much lower luminosities than Sw1644+57.
  
In Table~\ref{tab:properties} we include the TDE spectroscopic sub-classes as outlined by \citet{vanvelzen2020}, namely TDE-H, TDE-H+He, and TDE-He.  We find that all events in our sample classified as TDE-H exhibit delayed radio emission.  We also find that we have two detections of TDE-H+He events (AT2019teq, AT2020neh), but the majority of our non-detections are of the TDE-H+He class. We will explore possible connections between the radio emission properties and the multi-wavelength properties in a companion paper (Alexander et al.~in prep). 

Finally, it is worth noting that the TDEs with no radio detections may in fact turn on at a later time than the observations presented here. For example, ASASSN-14ae is the oldest TDE in our sample and its first detection was at $\approx 2300$ days.  Regular monitoring of the TDE population is crucial for determining the distribution of turn-on timescales; for example, in our own study, AT2018zr was not detected by two dedicated observations at 929 and 1218 days, but was then detected in a third observation at 1713 days.  VLASS data was not constraining for this source, and without our prior deep upper limits it would have been difficult to constrain the timescale at which radio emission turned on.

\begin{figure*}[t!] 
    \includegraphics[width=.45\columnwidth]{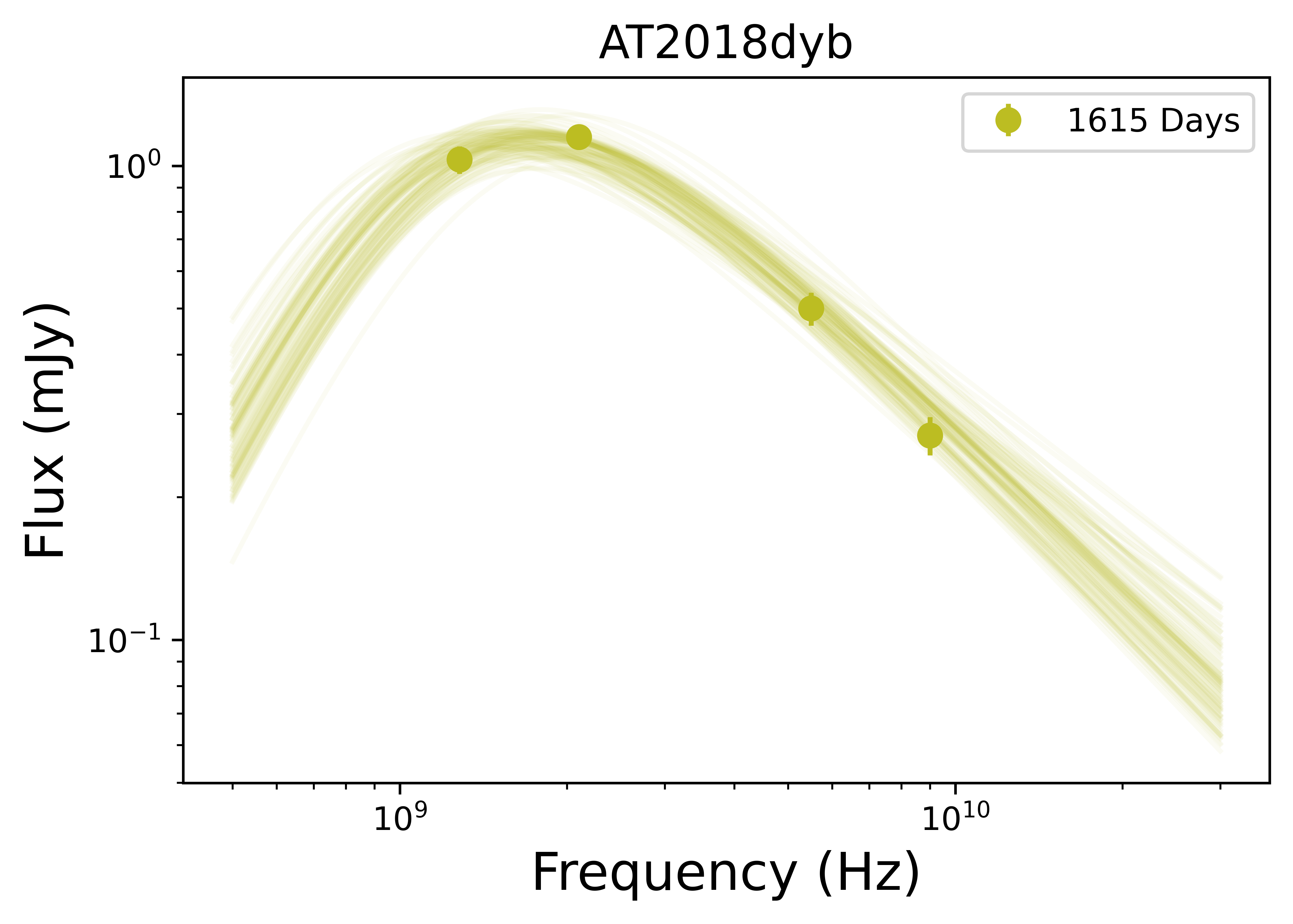}  \hfill
    \includegraphics[width=.45\columnwidth]{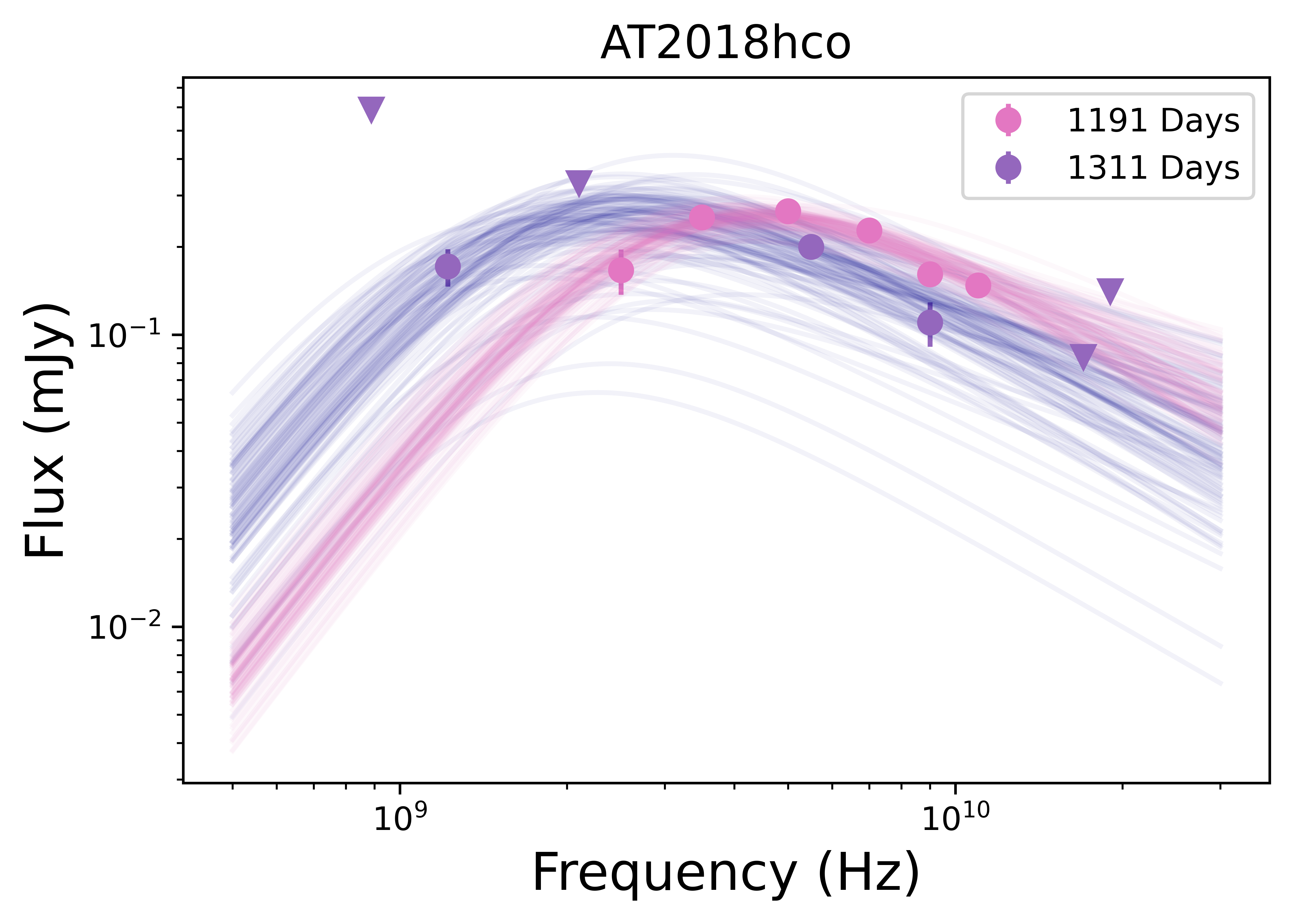}\hfill  
    \includegraphics[width=.45\columnwidth]{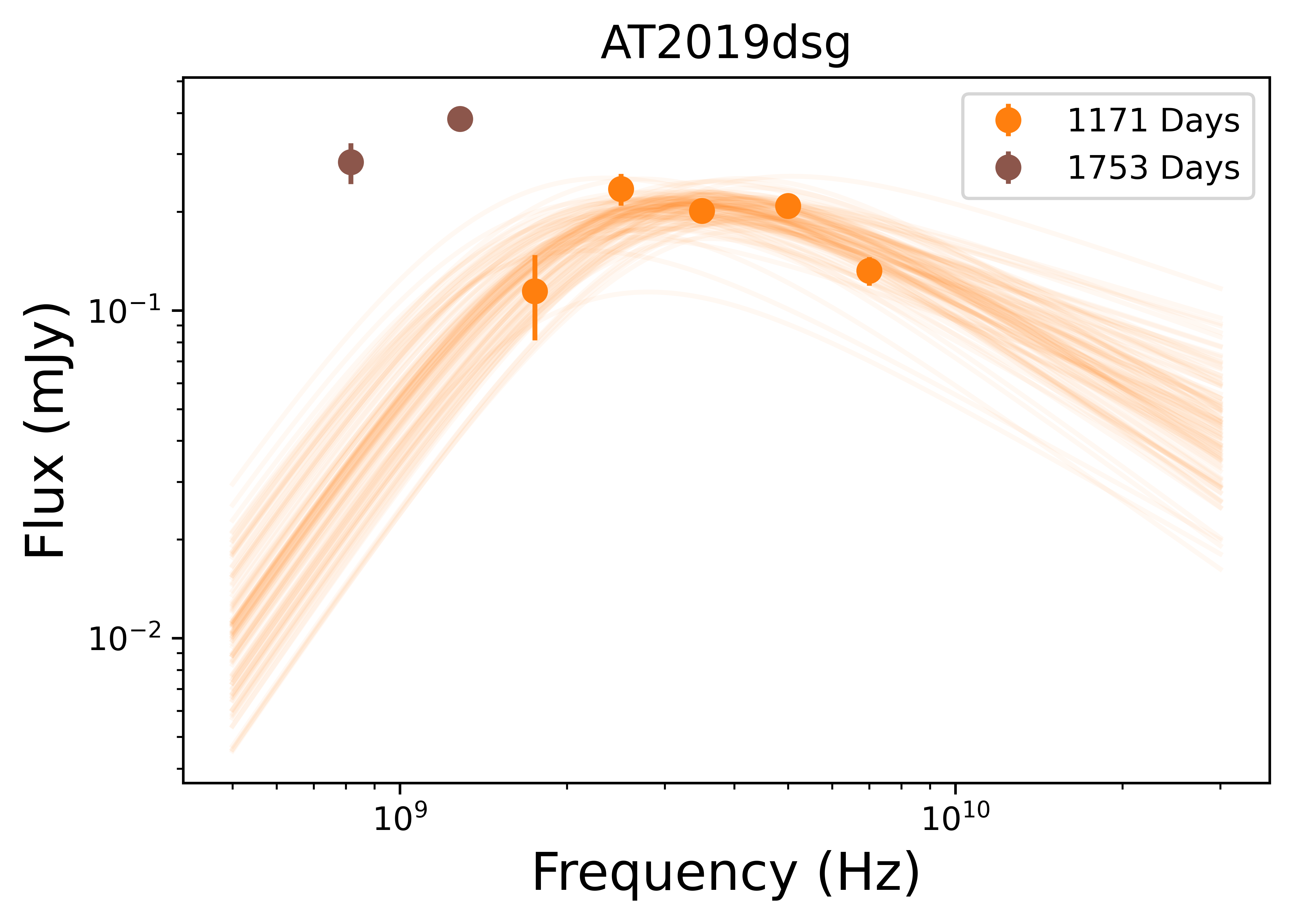}\hfill
    \includegraphics[width=.45\columnwidth]{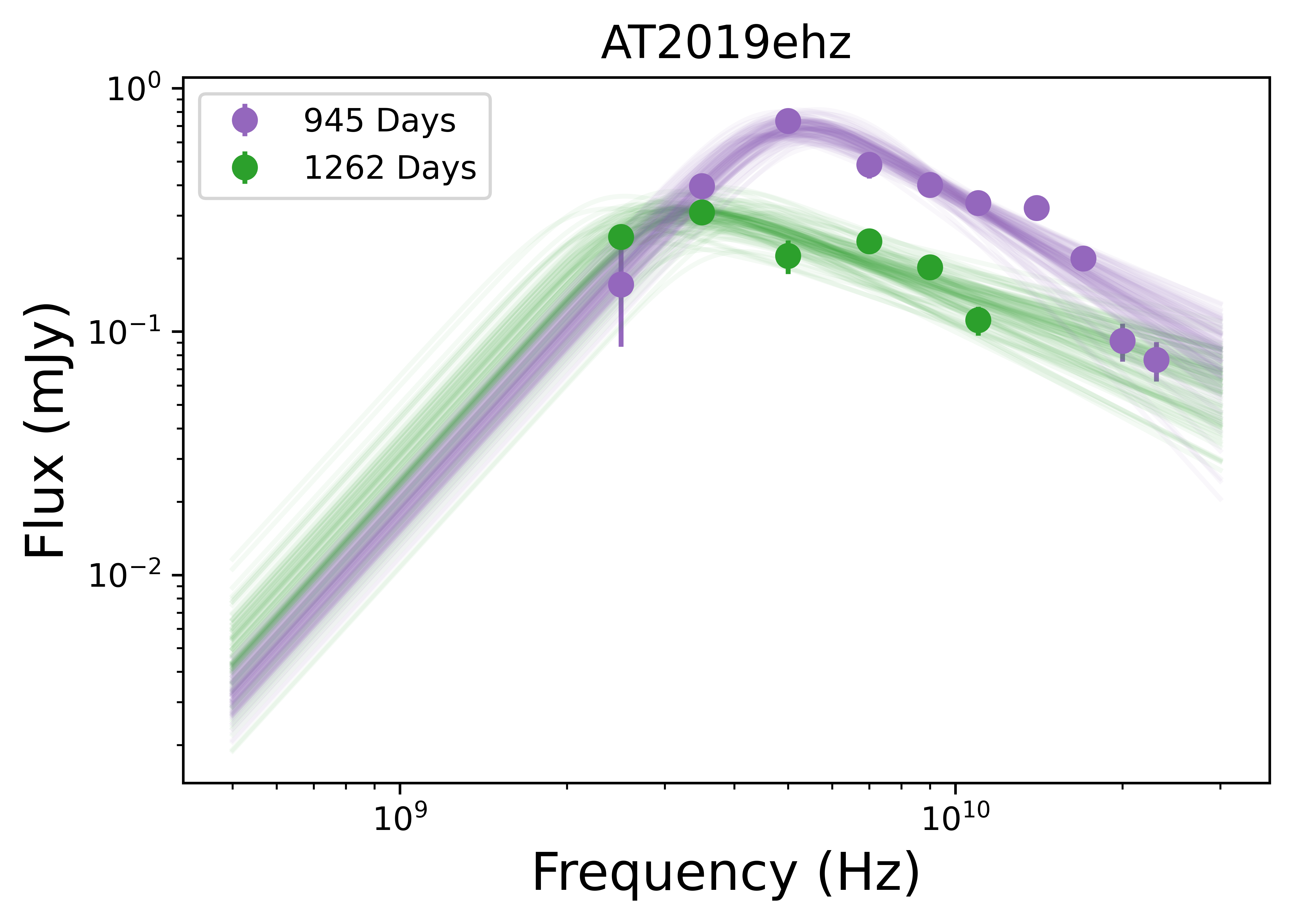}\hfill 
    \includegraphics[width=.45\columnwidth]{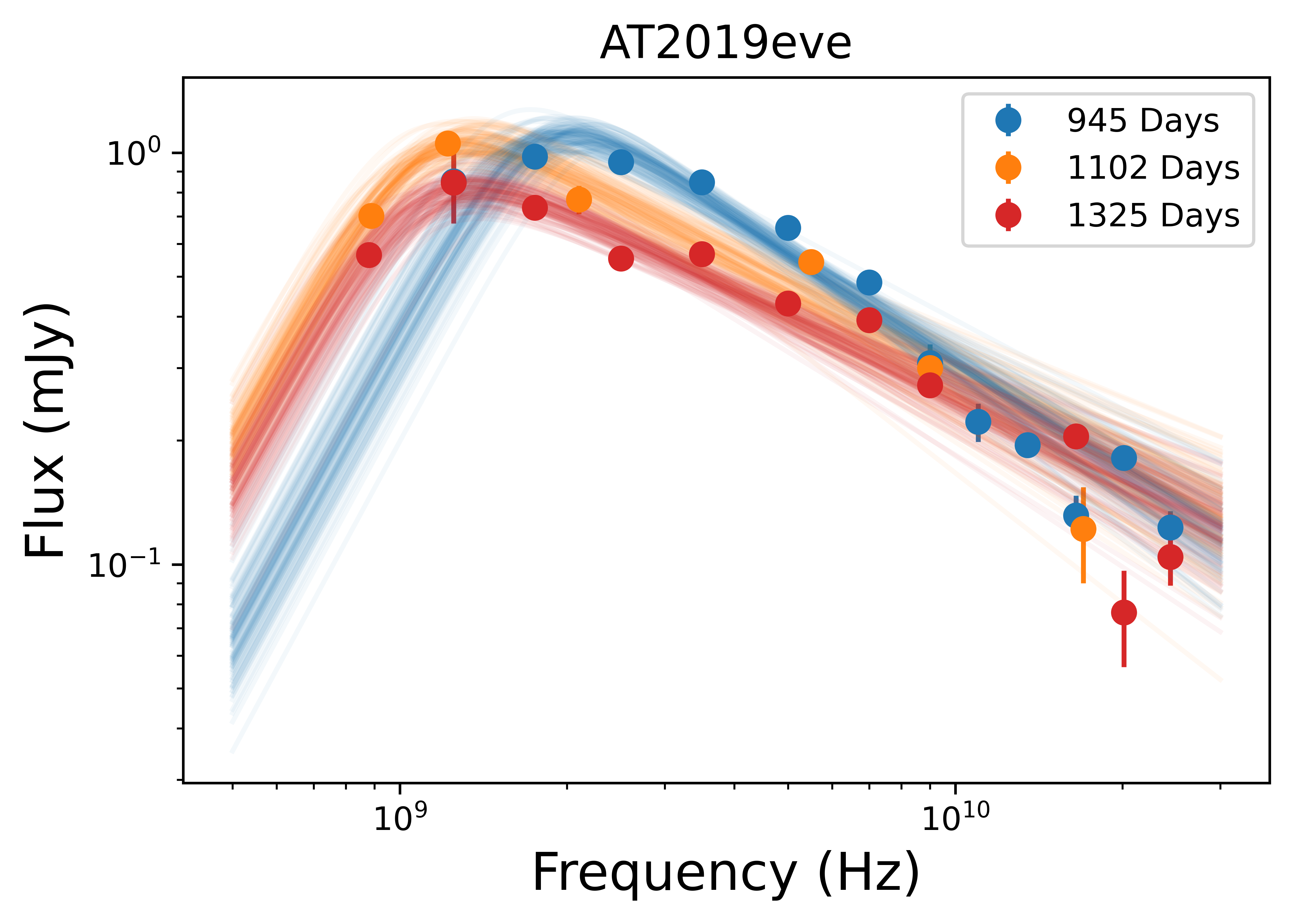}  %\hfill  
    \label{fig:sed}
    \caption{Radio spectral energy distributions (SEDs) for TDEs where $\nu_p$ is constrained.  We denote upper limits as triangles, and do not include non-constraining upper limits in these plots (they are available in Table~\ref{tab:obs}).  The lines are representative fits from our MCMC modeling of synchrotron self-absorbed spectra (\S\ref{sec:sed}).}
\end{figure*}

\begin{figure*}[t!]
    \includegraphics[width=.45\columnwidth]{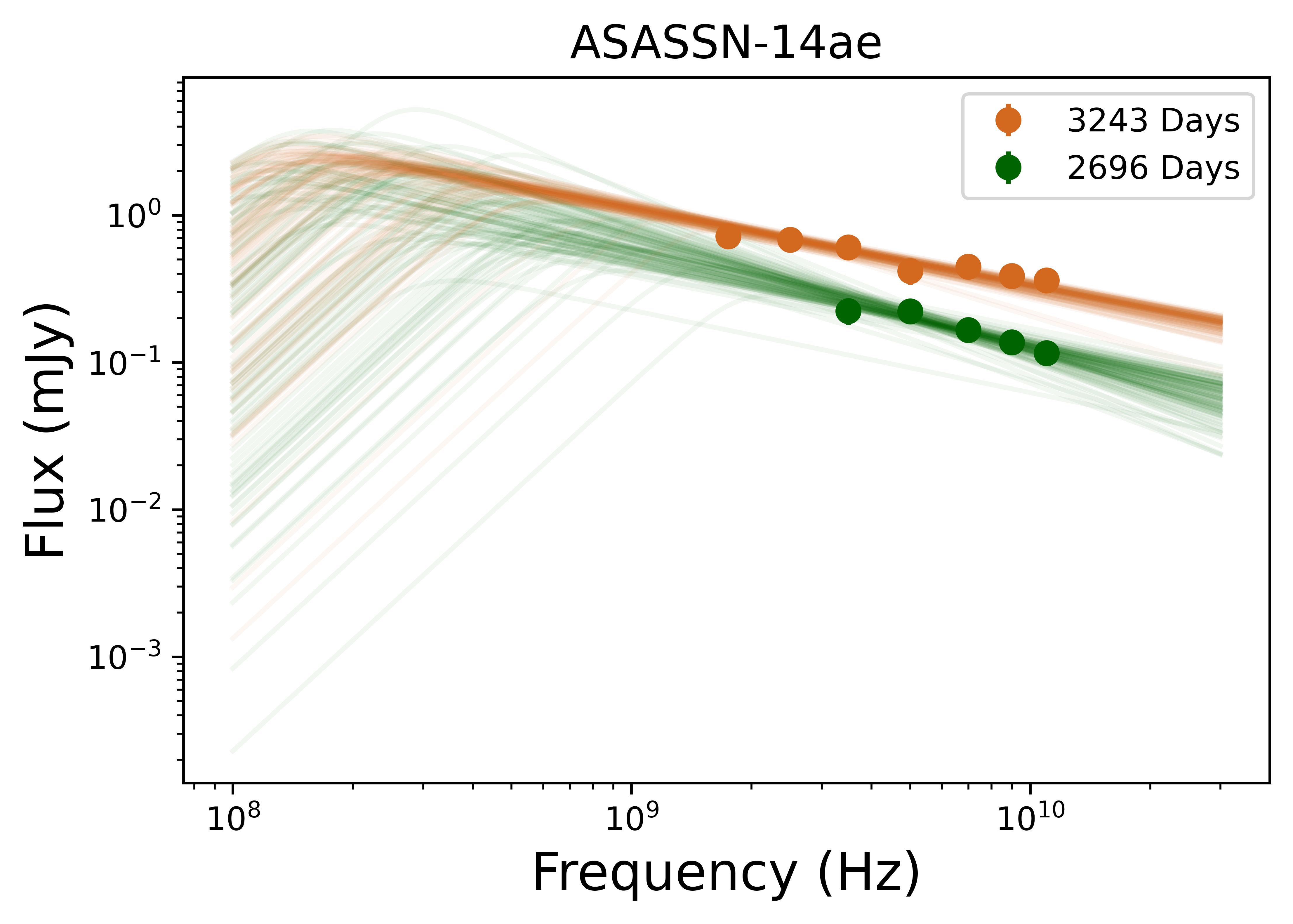}
    \includegraphics[width=.45\columnwidth]{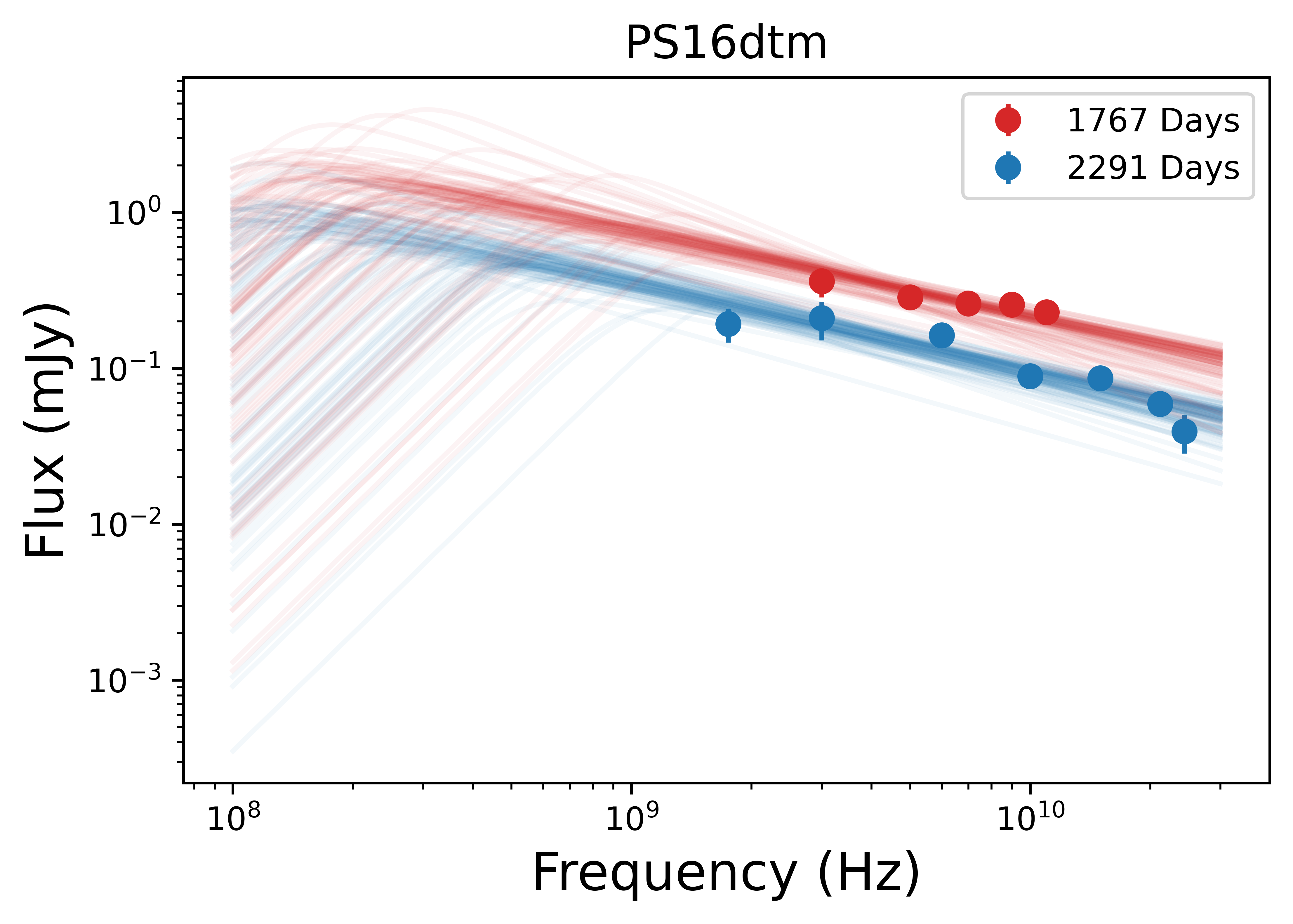}\hfill
    \includegraphics[width=.45\columnwidth]{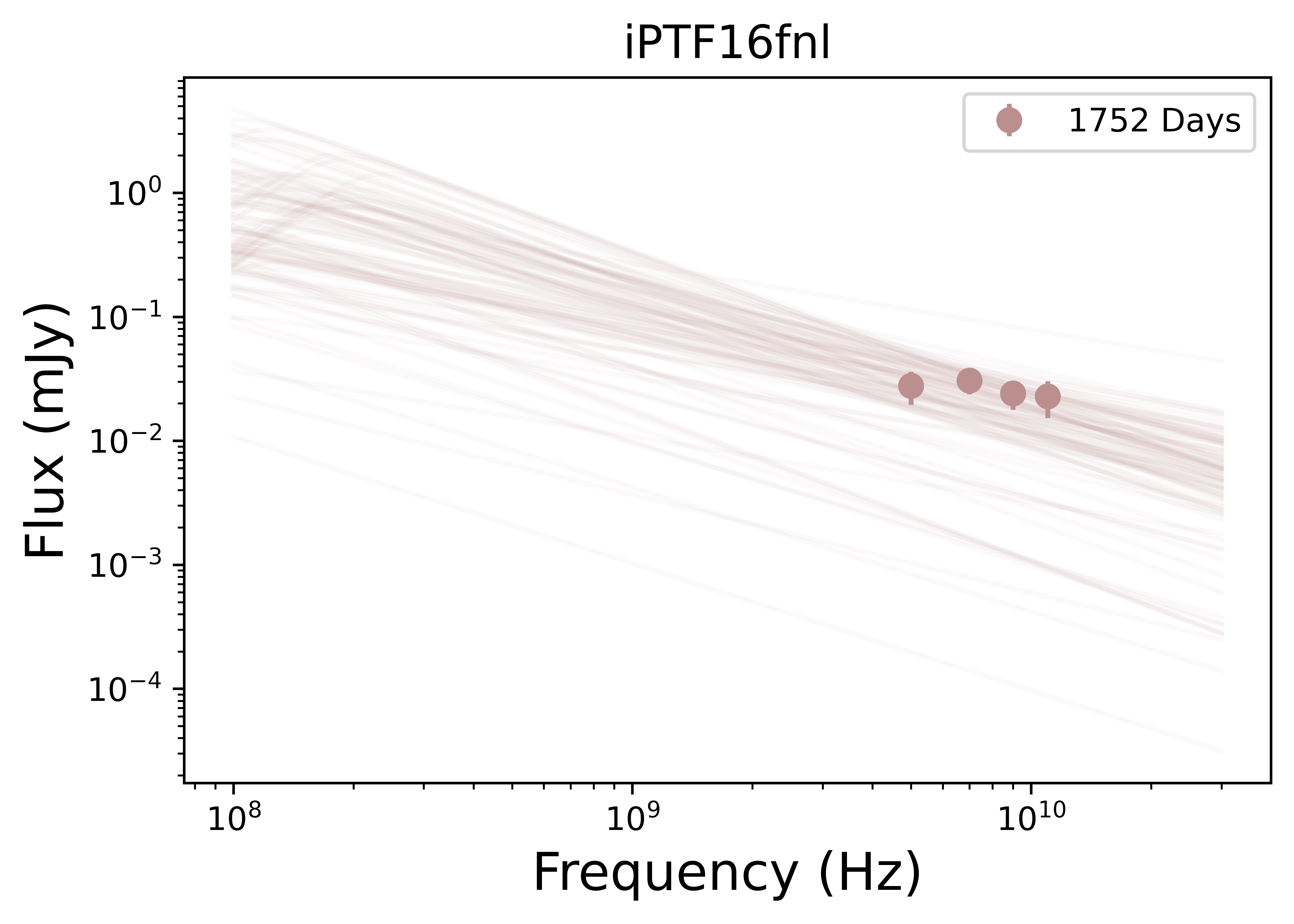}\hfill
    \includegraphics[width=.45\columnwidth]{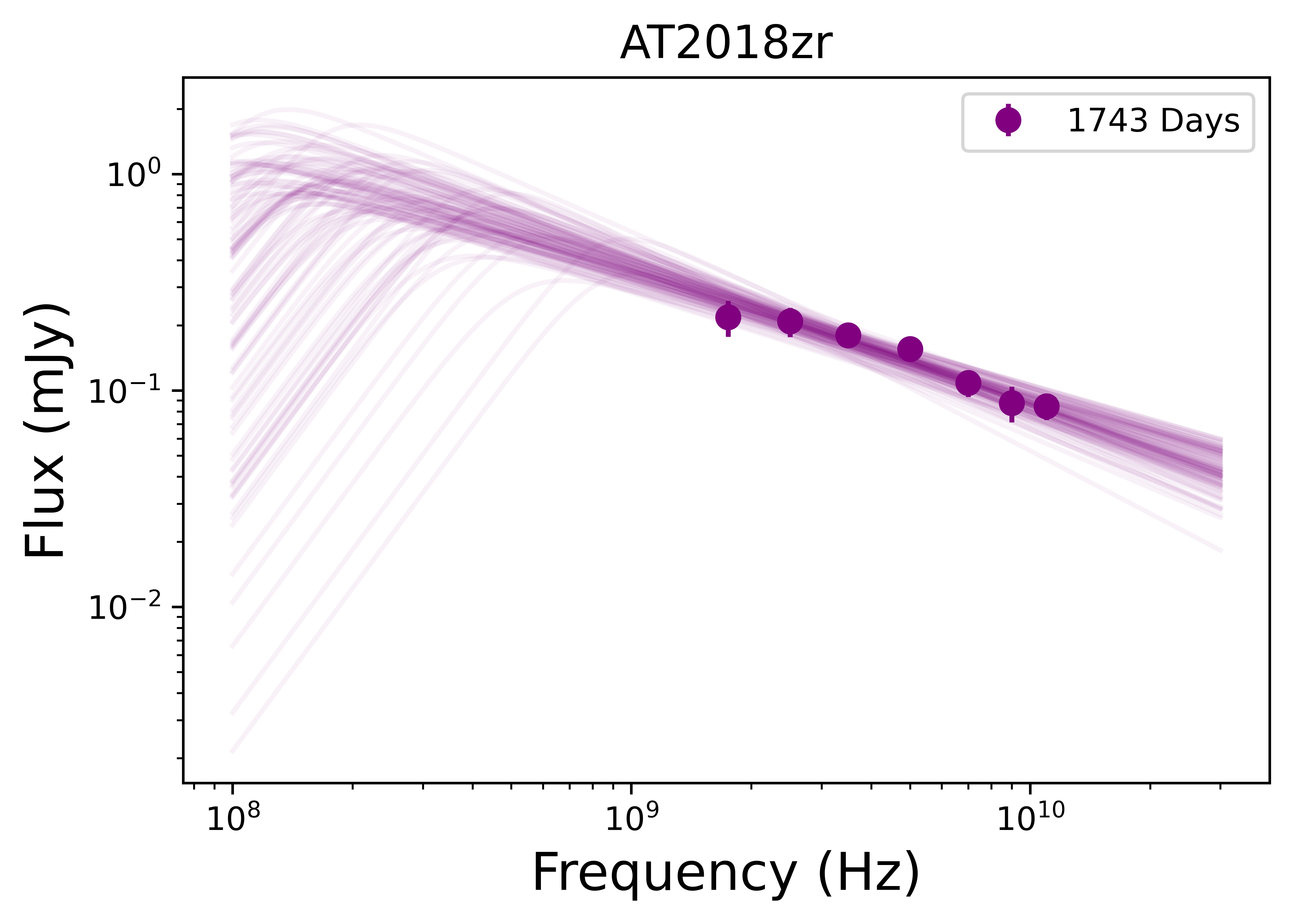}\hfill
    \includegraphics[width=.45\columnwidth]{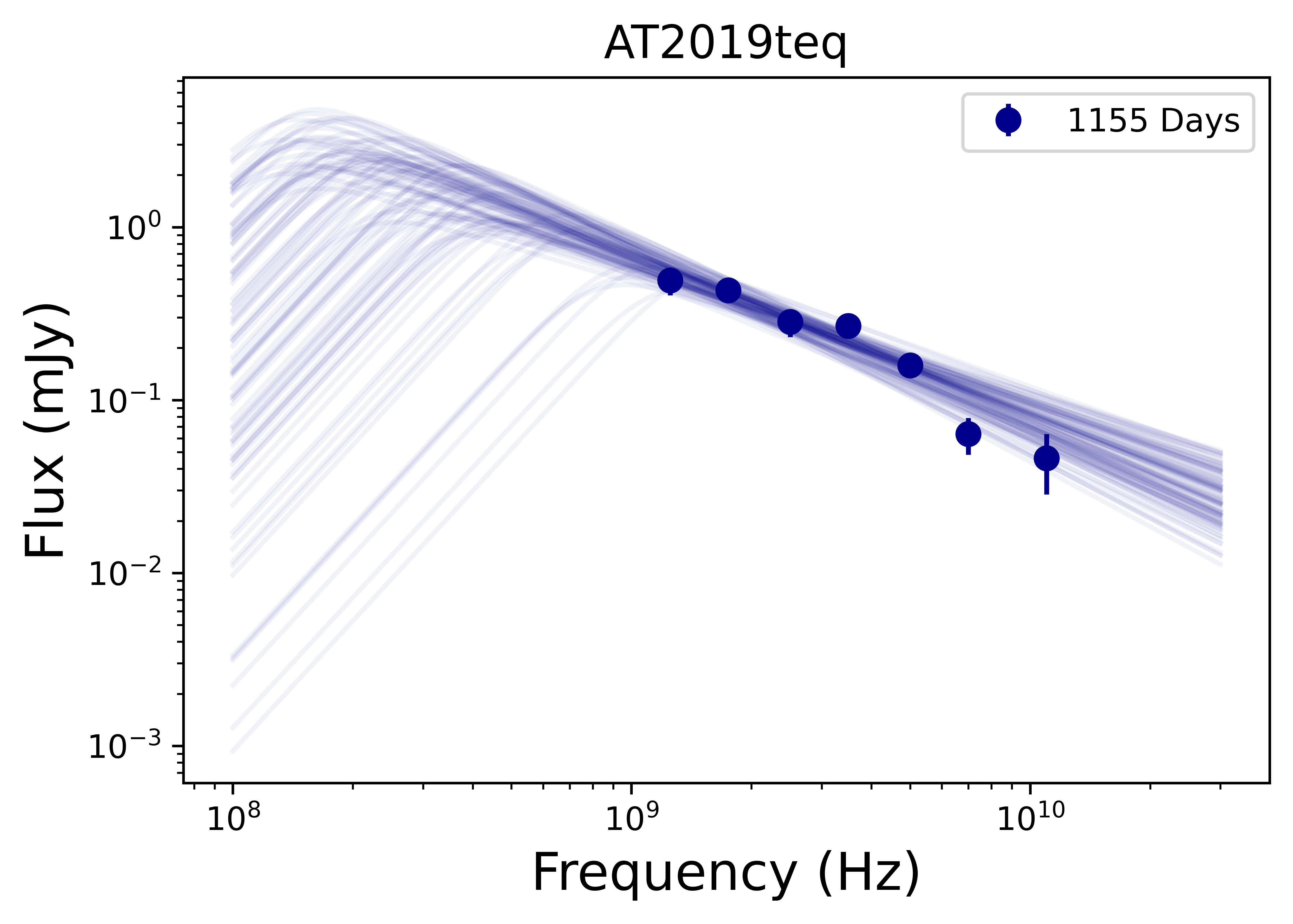} %\hfill
    \label{fig:sed-2}
    \caption{The same as Figure \ref{fig:sed}, but for TDEs where the SED is not constrained.}
\end{figure*}

\begin{deluxetable*}{lcccc}
\label{tab:sed}
\tablecolumns{5}
\tablecaption{Spectral Energy Distribution Parameters}
\tablehead{
	\colhead{TDE} &
	\colhead{$\delta t$} &
	\colhead{$F_{\nu,p}$} &
	\colhead{$\nu_p$} &
    \colhead{$p$} 
	\\
	\colhead{} &
    \colhead{(d)} &
    \colhead{(mJy)} &
    \colhead{(GHz)} &
    \colhead{} 
}
 \startdata
ASASSN-14ae$^*$ & 2696 & >0.329 & <2.5 &$2.25\substack{+0.24 \\ -0.16}$ \\
ASASSN-14ae$^*$ & 3243 & >0.720 & <1.75 &$2.05\substack{+0.10 \\ -0.04}$\\
iPTF16fnl$^*$ & 1752 & >0.028 & <5.0 &$2.69\substack{+0.54 \\ -0.48}$ \\
PS16dtm$^*$ & 1767 & >0.361 & <3.0 &$2.14\substack{+0.25 \\ -0.11}$\\
PS16dtm$^*$ & 2291 & >0.193 & <1.75 &$2.18\substack{+0.20 \\ -0.12}$\\
AT2018dyb & 1615 & $0.50\substack{+0.06 \\ -0.08}$ & $1.41\substack{+0.33 \\ -0.18}$ &$2.96\substack{+0.24 \\ -0.21}$\\
AT2018hco & 1191 & $0.26\substack{+0.02 \\ -0.01}$ & $4.37\substack{+0.38 \\ -0.38}$ &$3.03\substack{+0.32 \\ -0.48}$\\
AT2018hco & 1311 &  $0.25\substack{+0.06 \\ -0.05}$ & $2.75\substack{+0.56 \\ -0.52}$ &$2.84\substack{+0.13 \\ -0.14}$\\
AT2018zr$^*$ & 1743 & >0.218 & <1.75 &$2.26\substack{+0.20 \\ -0.17}$\\
AT2019dsg & 1171 & $0.21\substack{+0.02 \\ -0.02}$ & $3.31\substack{+0.58 \\ -0.58}$&$2.70\substack{+0.10 \\ -0.11}$ \\
AT2019ehz & 970 & $0.70\substack{+0.06 \\ -0.06}$ & $5.22\substack{+0.03 \\ -0.03}$&$3.63\substack{+0.40 \\ -0.34}$\\ % p
AT2019ehz & 1262 & $0.23\substack{+0.03 \\ -0.03}$ & $2.96\substack{+0.05 \\ -0.04}$&$2.55\substack{+0.45 \\ -0.33}$\\ % p
AT2019eve & 945 &  $1.12\substack{+0.06 \\ -0.06}$ & $2.00\substack{+0.14 \\ -0.14}$ &$2.75\substack{+0.14 \\ -0.17}$\\
AT2019eve & 1102 &  $ 1.06 \substack{+0.07 \\ -0.07}$ & $1.29\substack{+0.06 \\ -0.06}$& $2.37 \substack{+0.21 \\ -0.17}$ \\
AT2019eve & 1325 &  $ 0.82 \substack{+0.05 \\ -0.06}$ & $1.35 \substack{+0.09 \\ -0.06}$ & $2.27 \substack{+0.15 \\ -0.14}$\\
AT2019teq$^*$ & 1155& >0.492 & <1.25 & $2.97\substack{+0.38 \\ -0.32}$\\
 \enddata
 \tablecomments{$^*$ Indicates SEDs in which the model peak is at or near the lowest frequency data point, and the resulting values of $F_{\nu,p}$ and $\nu_p$ are lower and upper limits, respectively.  We have combined data for some SEDs taken with different telescopes at roughly the same time for broader frequency coverage, and note these instances in \S\ref{sec:indi-seds}. }
 \end{deluxetable*}

\newpage
\section{Spectral Energy Distribution Modeling and Analysis}
\label{sec:sed}

Beyond the light curve data discussed in the previous section, we have also obtained multi-frequency data for 10 TDEs with late-time radio emission \footnote{We also obtained SEDs for OGLE17aaj and AT2020pj, but the origin of the radio emission in these sources is ambiguous (\S\ref{sec:ambiguous}). We exclude these sources from our analysis, but provide the data for completeness in Table~\ref{tab:obs-not-used}.}, allowing us to model their spectral energy distributions (SEDs) and extract physical properties of the outflows. The SEDs are characteristic of self-absorbed synchrotron emission, with (in some cases) a well defined peak frequency ($\nu_p$) and peak flux density ($F_{\nu,p}$), and a spectral shape of $F_\nu\propto \nu^{5/2}$ below $\nu_p$ (Figure~\ref{fig:sed}). We chose this as the simplest explanation for the spectral shape, as other shapes (such as $F_\nu\propto \nu^{2}$) would require a relativistic outflow, and this does not seem to be the case for the majority of TDEs \citep{Alexander2020}.We note that in some other types of transients, such as supernovae \citep[e.g.,][]{Chandra2009}, shallow optically thick spectra have been observed and interpreted as inhomogenaities.  In this case, the peak frequency and flux translate into limits on the radius of the outflow. In several cases we find that the SED peak is below our lowest available frequency; we assume a self-absorbed synchrotron emission for these cases as well, and use the lowest-frequency data point as an upper limit on $\nu_p$ and a lower limit on $F_{\nu,p}$ (Figure~\ref{fig:sed-2}).  For 6 TDEs, we have 2 or 3 epochs of multi-frequency observations, while for 4 sources we have only a single epoch.

We fit the SEDs with the model of \citet{Granot2002}, developed for synchrotron emission from gamma-ray burst (GRB) afterglows, and previously applied to the radio emission from TDEs (e.g., \citealt{Zauderer2011,Cendes2021,Cendes2022}), using specifically the regime of $\nu_m\ll\nu_a$: 
\begin{equation}
\indent F_\nu = F_\nu (\nu_m) \Bigg[ \Big(\frac{\nu}{\nu_m}\Big)^2 e^{-s_4 (\nu/\nu_m)^{2/3}}+\Big(\frac{\nu}{\nu_m}\Big)^{5/2}\Bigg] 
\times \Bigg[1+\Big(\frac{\nu}{\nu_a}\Big)^{s_5 (\beta_2-\beta_3)}\Bigg]^{-1/s_5},
\label{eq:second-spec}
\end{equation}
where $\beta_2 = 5/2$, $\beta_3 = (1-p)/2$, $s_4 = 3.44p - 1.41$, and $s_5 = 1.47 - 0.21p$ \citep[as in ][]{Eftekhari2018,Cendes2021b}. Here, $p$ is the electron energy distribution power law index, $N(\gamma_e)\propto \gamma_e^{-p}$ for $\gamma_e\ge\gamma_m$, $\nu_m$ is the frequency corresponding to $\gamma_m$, $\nu_a$ is the synchrotron self-absorption frequency, and $F_\nu(\nu_m)$ is the flux normalization at $\nu=\nu_m$.

We determine the best fit parameters of the model --- $F_\nu(\nu_m)$, $\nu_a$, and $p$ --- using the Python Markov Chain Monte Carlo (MCMC) module \texttt{emcee} \citep{Foreman-Mackey2013}, assuming a Gaussian likelihood where the data have a log distribution for the parameters $F_{\nu}(\nu_m)$ and $\nu_a$, and a lower limit of $\nu_a > 0.1$ GHz (where the upper range for $\nu_a = 2-6$ GHz, based on the existing frequency for the individual SED).  For $p$ we use a uniform prior of $p=2-4.0$. We also include in the model a parameter that accounts for additional systematic uncertainty beyond the statistical uncertainty on the individual data points, $\sigma \lesssim 10\%$, which is this a fractional error added to each data point.  The posterior distributions are sampled using 100 MCMC chains, which were run for 3,000 steps, discarding the first 2,000 steps to ensure the samples have sufficiently converged by examining the sampler distribution. 

Our population has a range of SED values, with a broad range of values in p ($p \approx 2-3.6$). Further, in some cases with multiple SEDs (ASASSN-14ae, AT2019ehz, AT2019eve), we find the p value shows significant variation over several epochs.  Such a range and variation has been seen in prior radio TDEs \citep[e.g., ][]{Cendes2021,Goodwin2022}, and can be attributed to several reasons.  First, in the case of SEDs where $\nu_p$ is unconstrained, the paucity of data can lead to inconsistent values of p.  Our remaining variation in p is within ranges in other TDEs \citep{p1,Cendes2021b,Goodwin2022}.  We thus adopt the values for p for specific TDEs as outlined in Section \ref{sec:indi-seds} for subsequent analysis.

\subsection{Notes on Individual TDEs}
\label{sec:indi-seds}

The SED model fits are shown in Figures~\ref{fig:sed} and \ref{fig:sed-2} and provide a good match to the data, although in the cases in \ref{fig:sed-2} we do not constrain $\nu_a$.  In these cases, we adopt the lowest frequency data point as an upper limit, and the flux measured at that data point as a lower limit. The resulting model parameters are listed in Table~\ref{tab:sed}. Below we summarize key results for the individual TDEs.  

\begin{itemize}
    \item \textbf{ASASSN-14ae} has two SEDs obtained with the VLA at 2696 days and 3243 days.  In both cases, the SEDs peak below the lowest observed frequency.  We thus use the lowest frequency data point for each observation for the values of $F_{\nu,p}$ as a lower limit and $\nu_p$ as an upper limit.  We use a mean value of $p=2.15$ in the subsequent equipartition analysis.

    \item \textbf{PS16dtm} has two SEDs obtained with the VLA at 1767 and 2291 days.  In both cases the SEDs peak below the lowest observed data point, so we use the lowest data point for $\nu_p$ and $F_{\nu,p}$, and they are considered as upper and lower limits, respectively.  We use a mean value of $p=2.16$ in the subsequent equipartition analysis.
    
    \item \textbf{AT2018hco} has two SEDs at 1191 days (VLA) and 1311 days (combined ATCA on 2022 April 22 and MeerKAT on 2022 May 7).  We find that $F_{\nu,p}$ remained steady between the two observations, while $\nu_p$ decreased by about a factor of 1.6.  We use a mean value of $p=2.94$ in the subsequent equipartition analysis.
    
    \item \textbf{AT2019dsg} had a single SED at 1171 days (VLA), and a partial SED at 1753 days (MeerKAT). While the latter does not allow for a full SED fit, it clearly indicates a brightening by a factor of $\approx 4$ at low frequencies.  At 1171 days, we find $p=2.7$, consistent with the radio SEDs at earlier times \citep{Cendes2021b}.  However, while $F_{\nu,p}$ exhibits a continued decline, $\nu_p$ increases to $\approx 3.5$ GHz at 1171 days, while it was $\approx 1.7$ GHz at 561 days \citep{Cendes2021b}; this increase creates the rise at 6 GHz seen in Figure~\ref{fig:lumin-tde}.

    \item \textbf{AT2019ehz} had two SEDs at 970 and 1262 days obtained with the VLA.  We find that $F_{\nu,p}$ decreases by about a factor of 2 while $\nu_p$ decreases by about a factor of 1.5.  We also find a decline in the value of $p$ from about 3.6 to 2.6, and adopt a mean value of $p=3.1$ in the subsequent equipartition analysis. 

    \item \textbf{AT2019eve} has 3 SEDs at 945 (VLA), 1102 (combined ATCA on 2022 April 30 and MeerKAT on 2022 May 10), and 1325 (combined VLA on 2022 December 19 and MeerKAT on 2023 January 4) days. We note that $F_{\nu,p}$ remains steady at 945 and 1102 days, and then declines at 1325 days, while $\nu_p$ declines at 945 and 1102 days, but remains steady at 1325 days. We also find a decline in the value of $p$ from about 2.75 to 2.3, and we use a mean value of $p=2.46$ in the subsequent equipartition analysis. 

    \item \textbf{iPTF16fnl} has a single SED at 1752 days taken with the VLA. The SED is flat, with $\nu_p$ below our lowest data point, so we thus adopt these values for $\nu_p <5$ GHz and $F_{\nu,p} >0.028$.  We use $p=2.69$ in the subsequent equipartition analysis.
    
    \item \textbf{AT2018dyb} has a single SED at 1615 days (combined MeerKAT on 2022 December 11 and ATCA on 2023 January 22).  We find $F_{\nu,p}\approx 0.25$ mJy, $\nu_p \approx 3$ GHz, and $p=2.96$.
    
    \item \textbf{AT2018zr} has a single SED at 1743 days obtained with the VLA.  We find the SED peaks below our lowest observed frequency, and thus adopt $F_{\nu,p} > 0.329$, $\nu_p < 1.75$ GHz, and $p\approx 2.26$.
    
    \item \textbf{AT2019teq} has a single SED at 1155 days obtained with the VLA.  We find the SED peaks at the lowest observed frequency, indicating that the values of $\nu_p$ and $F_{\nu,p}$ may be considered as limits, and thus adopt $F_{\nu,p} > 0.492$ mJy, $\nu_p < 1.25$ GHz, and $p\approx 2.97$.  

\end{itemize}

\subsection{Equipartition Analysis}
\label{sec:equi}

\begin{deluxetable*}{lllllllll}
\label{tab:equi}
\tablecolumns{10}
\tablecaption{Equipartition Model Parameters}
\tablehead{
	\colhead{Object} &
	\colhead{$\delta t$} &
	\colhead{log($R_{\rm eq}$)} &
	\colhead{log($E_{\rm eq}$)} &
	\colhead{log($B$)} &
	\colhead{log($N_e$)} &
	\colhead{log($n_{\rm ext}$)} &
	\colhead{$\beta$} &
	\colhead{log($\nu_c$)} 
	\\
	\colhead{} &
    \colhead{(d)} &
    \colhead{(cm)} &
    \colhead{(erg)} &
    \colhead{(G)} &
    \colhead{} &
    \colhead{(cm$^{-3}$)} &
    \colhead{} &
    \colhead{(Hz)} 
}
\startdata
ASASSN-14ae$^*$ & 2696 & >16.61 & >47.70 & <$-0.54$ & >53.02 & <2.54 & >0.006 & >9.10 \\
ASASSN-14ae$^*$ & 3423 & >16.92 & >48.26 & <$-0.72$ & >53.59 & <2.16 & >0.010 & >9.15 \\
iPTF16fnl$^*$ & 1752 & >15.37 & >44.79 & <$-0.13$ & >50.55 & <3.79 & >0.0005 & >8.24 \\
PS16dtm$^*$ & 1767 & >16.77 & >48.10 & <$-0.58$ & >52.68 & <1.72 & >0.014 & >9.59 \\
PS16dtm & 2291 & >16.86 & >47.99 & <$-0.77$ & >52.57 & <1.32 & >0.013 & >9.96 \\
AT2018dyb & 1615 &$16.53\substack{+0.05 \\ -0.03}$ &$46.90\substack{+0.52 \\ -0.29}$&$-0.83\substack{+0.23 \\ -0.12}$&$52.54\substack{+0.08 \\ -0.12}$&$2.23\substack{+0.13 \\ -0.20}$&$ 0.008\substack{+0.002 \\ -0.002}$& $ 10.42\substack{+0.15 \\ -0.11}$ \\
AT2018hco & 1191 &$16.55\substack{+0.10 \\ -0.07}$&$47.76\substack{+1.06 \\ -0.63}$&$-0.43\substack{+0.39 \\ -0.24}$&$53.10\substack{+0.14 \\ -0.13}$&$2.78\substack{+0.31 \\ -0.15}$&$0.012\substack{+0.003 \\ -0.002}$& 9.48$\substack{+1.24 \\ -1.42}$\\
AT2018hco & 1311 &$16.60\substack{+0.17 \\ -0.12}$&$47.58\substack{+1.21 \\ -0.90}$&$-0.59\substack{+0.39 \\ -0.30}$&$52.91\substack{+0.21 \\ -0.13}$&$2.44\substack{+0.39 \\ -0.42}$&$0.012\substack{+0.01 \\ -0.01}$& $9.88\substack{+1.14 \\ -1.40}$\\
AT2018zr$^*$ & 1743 & >16.84 & >47.91 & <$-0.78$ & >52.51 & <1.34 & >0.016 & >10.21\\
AT2019dsg & 1171 &$16.44\substack{+0.08 \\ -0.07}$&$47.29\substack{+0.32 \\ -0.30}$&$-0.50\substack{+0.13 \\ -0.12}$&$52.80\substack{+0.10 \\ -0.09}$&$2.82\substack{+0.17 \\ -0.17}$&$0.009\substack{+0.002 \\ -0.002}$& $9.71\substack{+0.03 \\ -0.04}$\\
AT2019ehz & 970 &$16.61\substack{+0.03 \\ -0.03}$ &$47.95\substack{+0.04 \\ -0.04}$&$-0.41\substack{+0.02 \\ -0.02}$&$53.37\substack{+0.05 \\ -0.05}$&$2.87\substack{+0.05 \\ -0.06}$&0.017$ \substack{+0.001 \\ -0.001}$& $ 9.62\substack{+0.06 \\ -0.06}$ \\
AT2019ehz & 1262 &$16.63\substack{+0.05 \\ -0.05}$ &$47.68\substack{+0.07 \\ -0.07}$&$-0.59\substack{+0.05 \\ -0.05}$&$53.04\substack{+0.09 \\ -0.07}$&$2.48\substack{+0.08 \\ -0.11}$&0.014$ \substack{+0.002 \\ -0.002}$& $ 9.91\substack{+0.13 \\ -0.16}$ \\
AT2019eve & 945 &$17.18\substack{+0.03 \\ -0.03}$&$48.86\substack{+0.04 \\ -0.04}$&$-0.81\substack{+0.03 \\ -0.03}$&$54.38\substack{+0.25 \\ -0.24}$&$2.19\substack{+0.28 \\ -0.26}$&$0.061\substack{+0.003 \\ -0.004}$ & $10.83\substack{+0.08 \\ -0.10}$ \\
AT2019eve & 1102 &$17.34\substack{+0.02 \\ -0.02}$&$49.00\substack{+0.04 \\ -0.04}$&$-0.99\substack{+0.03 \\ -0.03}$&$54.02\substack{+0.29 \\ -0.32}$&$1.35\substack{+0.30 \\ -0.31}$&$0.075\substack{+0.004 \\ -0.004}$& $11.24\substack{+0.08 \\ -0.08}$ \\
AT2019eve & 1325 &$17.28\substack{+0.03 \\ -0.03}$&$48.84\substack{+0.05 \\ -0.05}$&$-0.98\substack{+0.03 \\ -0.03}$&$53.66\substack{+0.23 \\ -0.25}$&$1.16\substack{+0.22 \\ -0.24}$&$0.056\substack{+0.004 \\ -0.004}$ & $11.04\substack{+0.08 \\ -0.08}$ \\
AT2019teq$^*$ & 1155 & >17.23 & >48.65 & <$-0.99$ & >53.31 & <0.97 & >0.06 & >11.21 \\
\enddata
\tablecomments{Values in this table are calculated using an outflow launch time based on the optical discovery date, with errors are propagated from the $\nu_p$, $F_{\nu,p}$ errors in Table \ref{tab:sed}}.
$^*$ Indicates an unconstrained SED, where the resulting values can be considered as limits.\\
\end{deluxetable*}

Using the inferred values of $\nu_p$, $F_{\nu,p}$ and $p$ (Table~\ref{tab:sed}), we can now derive the physical properties of the outflows and ambient medium using an equipartition analysis. We assume the conservative case of a non-relativistic spherical outflow; in \S\ref{sec:jet} we demonstrate that none of the TDEs are in the regime where an off-axis relativistic jet interpretation can be supported  (i.e., using the criterion in \citealt{Matsumoto2023}). In the non-relativistic spherical case the radius and kinetic energy are given by \citep[see Equations 27 and 28 in][]{Duran2013}:

\begin{multline} 
R_{\rm eq}\approx (1\times10^{17}\,\, \textrm{cm})\times(21.8\times525^{p-1})^{\frac{1}{13+2p}} \chi_{e}^{\frac{2-p}{13+2p}}F_{p,\rm mJy}^{\frac{6+p}{13+2p}} d_{L,28}^{\frac{2(p+6)}{13+2p}} \nu_{p,10}^{-1}(1+z)^{\frac{-(19+3p)}{13+2p}}
 f_A^{\frac{-(5+p)}{13+2p}} f_V^{\frac{-1}{13+2p}} \\
\times 4^{\frac{1}{13+2p}}  \epsilon^{1/17}\xi^{\frac{1}{13+2p}},
\label{eq:rad}
\end{multline}
\begin{multline} 
E_{\rm eq} \approx (1.3\times10^{48}\,\, \textrm{erg})\times [21.8^{\frac{-2(p+1)}{13+2p}}][525^{p-1}\times \chi_e^{2-p}]^{\frac{11}{13+2p}} F_{p,\rm mJy}^{\frac{14+3p}{13+2p}} d_{L,28}^{\frac{2(3p+14)}{13+2p}} \nu_{p,10}^{-1} (1+z)^{\frac{-(27+5p)}{13+2p}}\\ 
\times f_A^{\frac{-(3(p+1)}{13+2p})} f_V^{\frac{2(p+1)}{13+2p}} 4^{\frac{11}{13+2p}} \xi^{\frac{11}{13+2p}} [(11/17)\epsilon^{-6/17}+(6/17)\epsilon^{11/17}],
\label{eq:energy}
\end{multline}
where $d_L$ is the luminosity distance; $z$ is the redshift; $f_A=1$ and $f_V= \frac{4}{3}\times (1-0.9^3)=0.36$ are the area and volume filling factors, respectively, where we assume that the emitting region is a shell of thickness $0.1R_{\rm eq}$; and $\gamma_m={\rm max}[\chi_e(\Gamma-1),2]$ is the minimum Lorentz factor as relevant for non-relativistic sources, and $\chi_e= (p-2)/(p-1)\epsilon_e(m_p/m_e)$, where $m_p$ and $m_e$ are the proton and electron masses, respectively. \citep{Duran2013}. The factors of $4^{\frac{1}{13+2p}}$ and $4^{\frac{11}{13+2p}}$ for the radius and energy, respectively, arise from corrections to the isotropic number of radiating electrons ($N_{e,\rm iso}$) in the non-relativistic case. We further assume that the fraction of post-shock energy in relativistic electrons is $\epsilon_e=0.1$, which leads to correction factors of $\xi^{\frac{1}{13+2p}}$ and $\xi^{\frac{11}{13+2p}}$ in $R_{\rm eq}$ and $E_{\rm eq}$, respectively, with $\xi = 1 + \epsilon_e^{-1}\approx 11$. Finally, we parameterize any deviation from equipartition with a correction factor $\epsilon = (11/6)(\epsilon_B/\epsilon_e)$, where $\epsilon_B$ = 0.1 is the fraction of post-shock energy in magnetic fields. 

Using $R_{\rm eq}$ we can also determine additional parameters of the outflow and environment \citep{Duran2013}: the magnetic field strength ($B$), the Lorentz factor of electrons radiating at $\nu_a$ ($\gamma_a$), and the number of radiating electrons ($N_e$):
\begin{equation}
B \approx (1.3\times10^{-2}\, \textrm{G})\times  F_{p,\rm mJy}^{-2} d_{L,28}^{-4} (1+z)^{7}  \nu_{p,10}^5 f_A^{2} R_{\rm eq,17}^{4}
\label{eq:mag}
\end{equation}
\begin{equation}
\gamma_a \approx 525\times F_{p,\rm mJy} d_{L,28}^2 \nu_{p,10}^{-2} (1+z)^{-3} \frac{1}{f_A R_{\rm eq,17}^{2}}.
\label{eq:gamma}
\end{equation}
\begin{equation}
N_{e} \approx (4\times10^{54})\times F_{p,\rm mJy}^{3} d_{L,28}^{6} (1+z)^{-8} f_A^{-2} R_{\rm eq,17}^{-4} \nu_{p,10}^{-5} \times (\gamma_{a}/\gamma_{m})^{(p-1)},
\label{eq:dens}
\end{equation}
We note an additional factor of 4 and a correction factor of $(\gamma_{a}/\gamma_{m})^{p-1}$ are added to $N_e$ for the non-relativistic regime (Barniol Duran, private communication). We determine the ambient density assuming a strong shock and an ideal monoatomic gas as $n_{\rm ext} = N_e/4V$, where the factor of 4 is due to the shock jump conditions and $V$ is the volume of the emitting region as defined above.

Finally, the synchrotron cooling frequency, $\nu_c$, is given by \citep{Sari1998}:
\begin{equation} 
\nu_c\approx 2.25\times 10^{14} B^{-3} \Gamma^{-1} t_d^{-2}\,\,{\rm Hz}.
\label{eq:cooling}
\end{equation}
where $t_d$ refers to the age of the system, and here $\Gamma\approx 1$.  The inferred values for $\nu_c$ are listed in Table~\ref{tab:equi}.  In several cases (e.g., AT2018hco, AT2019ehz) the value of $\nu_c$ is within the frequency range of our data, but we do not observe a cooling break.  This may be indicative of an incorrect value for $t_d$ in Equation~\ref{eq:cooling} due to a delayed outflow, as we discuss in \S\ref{sec:outflow-date}. Another possibility is that the value of $\epsilon_B$ is lower than our fiducial assumption of $0.1$, and hence the outflow may not be in equipartition. Although such deviations have been measured in TDEs in the past \citep{Eftekhari2018,Cendes2021b}, given the lack of evidence for such a deviation in our data we choose to conservatively assume no deviation is present. We note that our energy values are thus a lower limit and would be higher if a deviation from equipartition is assumed.

The inferred equipartition values for the 9 new TDEs with SEDs reported in this paper, and for the TDE with a distinct late-time component (AT2019dsg), and iPTF16fnl, are listed in Table~\ref{tab:equi}. Using the inferred values of $R_{\rm eq}$ we calculate the mean expansion velocity at each epoch, $\beta=v/c$, assuming that the outflow was launched at the time of optical discovery.

\subsection{Estimated Outflow Launch Timescales}
\label{sec:outflow-date}

Using the time of optical discovery as an initial estimate for the outflow launch date, the inferred velocities are in the range $\beta\approx 0.008-0.07$ (Table~\ref{tab:equi}), with the lowest velocities inferred for the older TDEs in our sample.  However, for TDEs that have more than one SED where we directly measure the peak (AT2019eve, AT2019ehz, and AT2018hco), we find higher values for $\beta$ by comparing radius measurements between SED epochs than what we infer from individual epochs with the assumption that the outflow began at the time of optical discovery.  This indicates that the assumption of a launch date that roughly coincides with the optical discovery is incorrect, as also indicated by non-detections at earlier times.

Thus, we estimate the launch dates of the outflows following the method in \citet{Cendes2022}; namely, we fit the radius evolution with a linear trend and determine the timescale at which $R=0$.  This analysis requires multiple SEDs, and so we only carry it out for 5 TDEs (ASASSN-14ae, PS16dtm, AT2018hco, AT2019ehz, AT2019eve). Since all of these TDEs also have an initial single-frequency (6 GHz) detection before an SED was obtained, we extract a rough radius estimate from this first detection by estimating the value of $\nu_p$ at that time from its later evolution, and calibrating $F_{\nu,p}$ to match the observed 6 GHz flux density. In the case of AT2019eve, the third (latest) SED exhibits a decrease in $F_{\nu,p}$ while $\nu_p$ remains relatively steady, resulting in a \textit{decrease} in the inferred radius compared to $\approx 200$ days earlier.  As the first two measured radii for AT2019eve show an increase, and the third SED occurred during a continued decrease of the light curve, we conclude it is more likely to be related to a change in the outflow's structure related to deceleration, which would not be useful for estimating the launch time of the outflow, and thus do not include this last data point in our calculation.  For ASASSN-14ae and PS16dtm, where the peak of emission appears to be below our lowest data point, we extract a rough radius estimate based off the lowest data point directly measured ($\nu_p = 1.75$ GHz in both cases), and assume $\nu_p$ is fixed at this value throughout our observation.

 \begin{figure*}
    \includegraphics[width=.32\columnwidth]{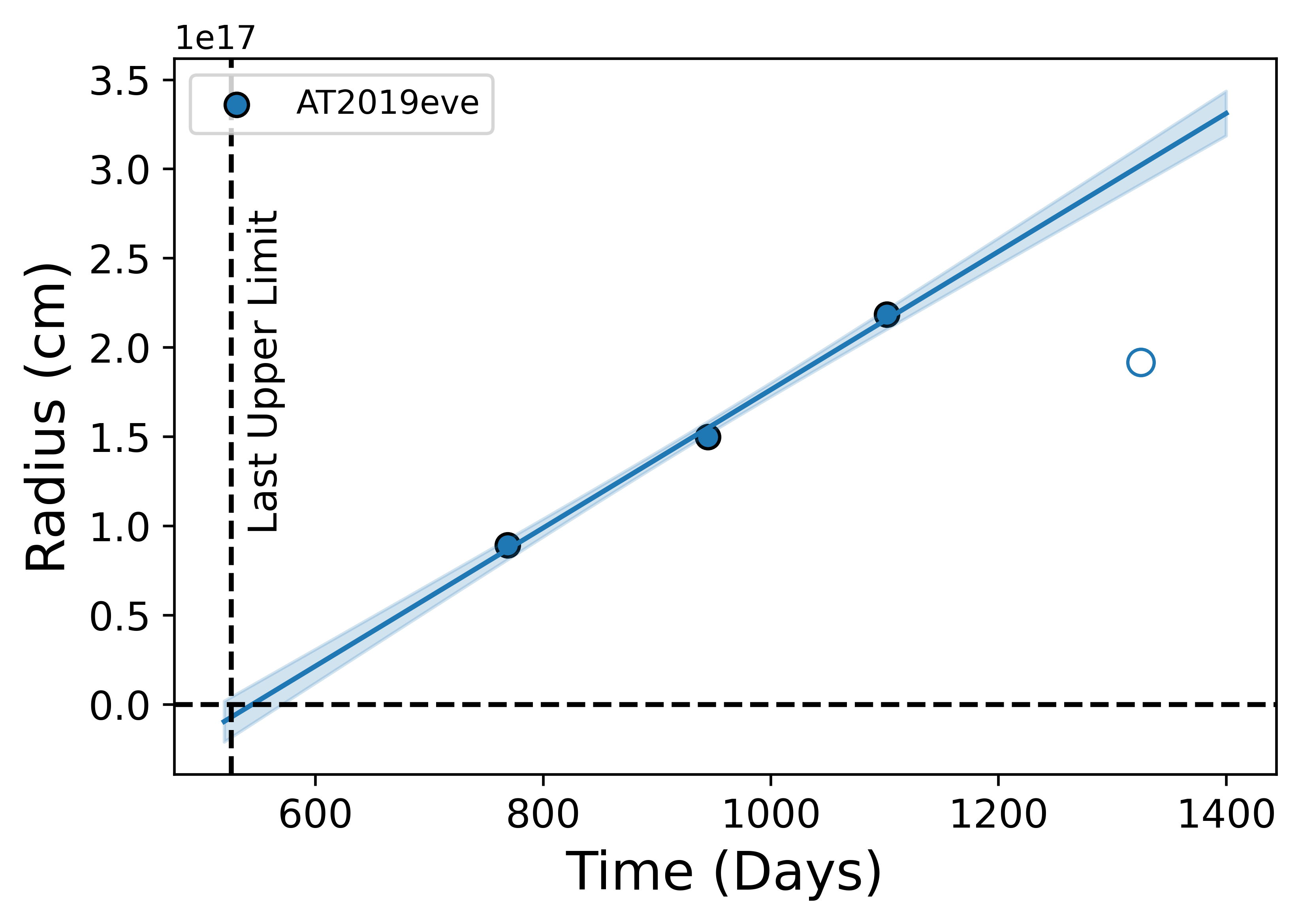}  
    \includegraphics[width=.32\columnwidth]{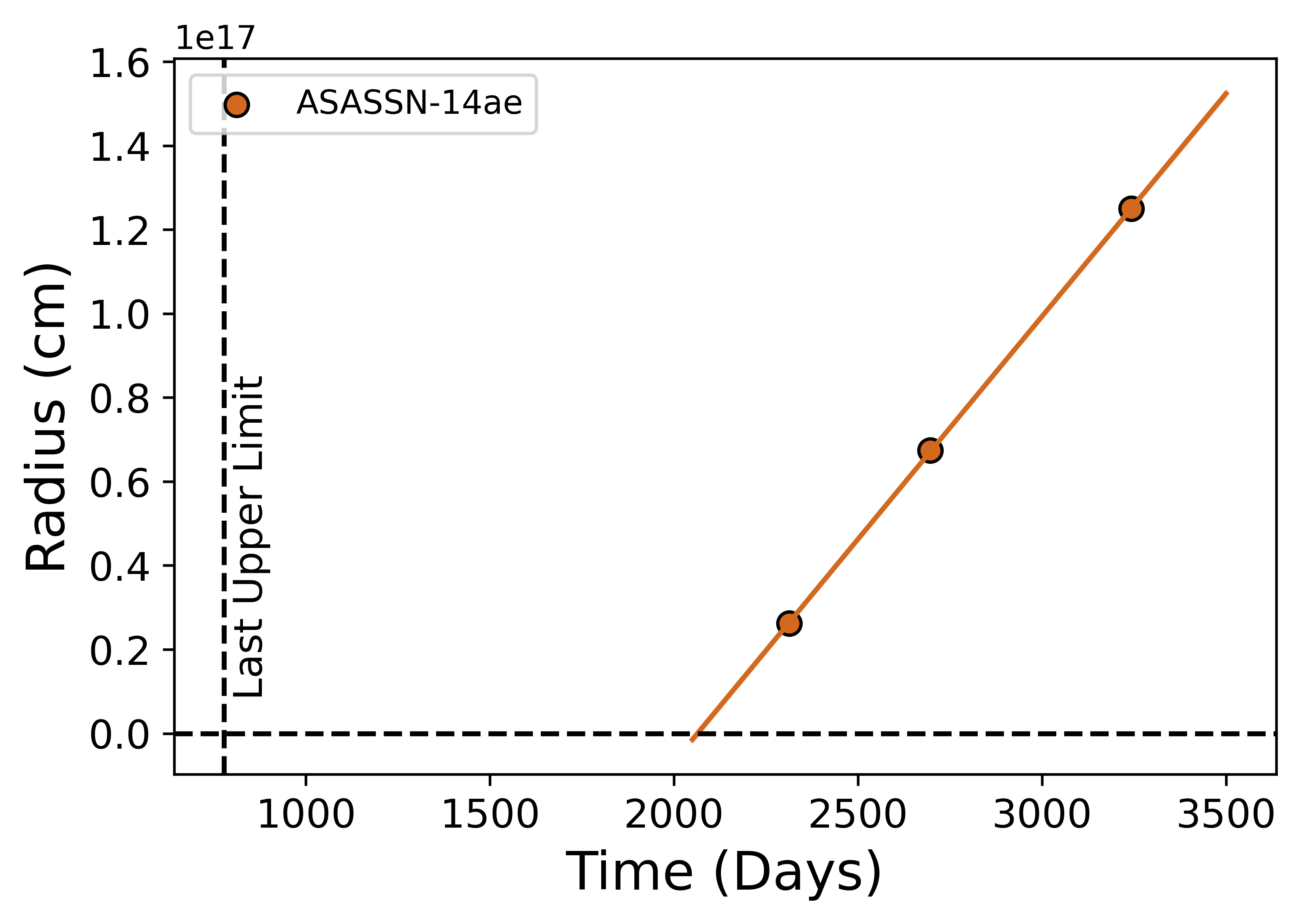}
    \includegraphics[width=.32\columnwidth]{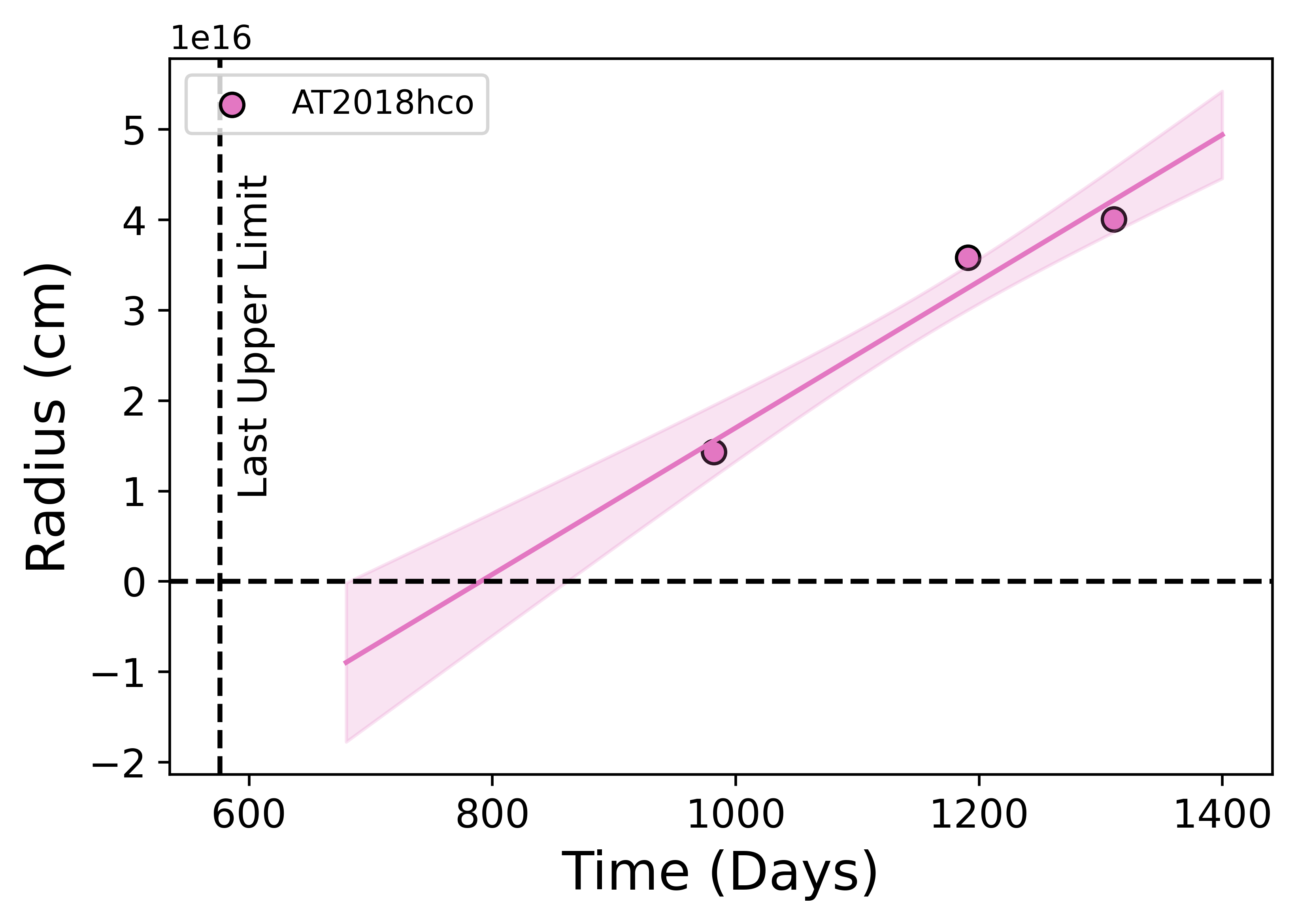}
    \includegraphics[width=.32\columnwidth]{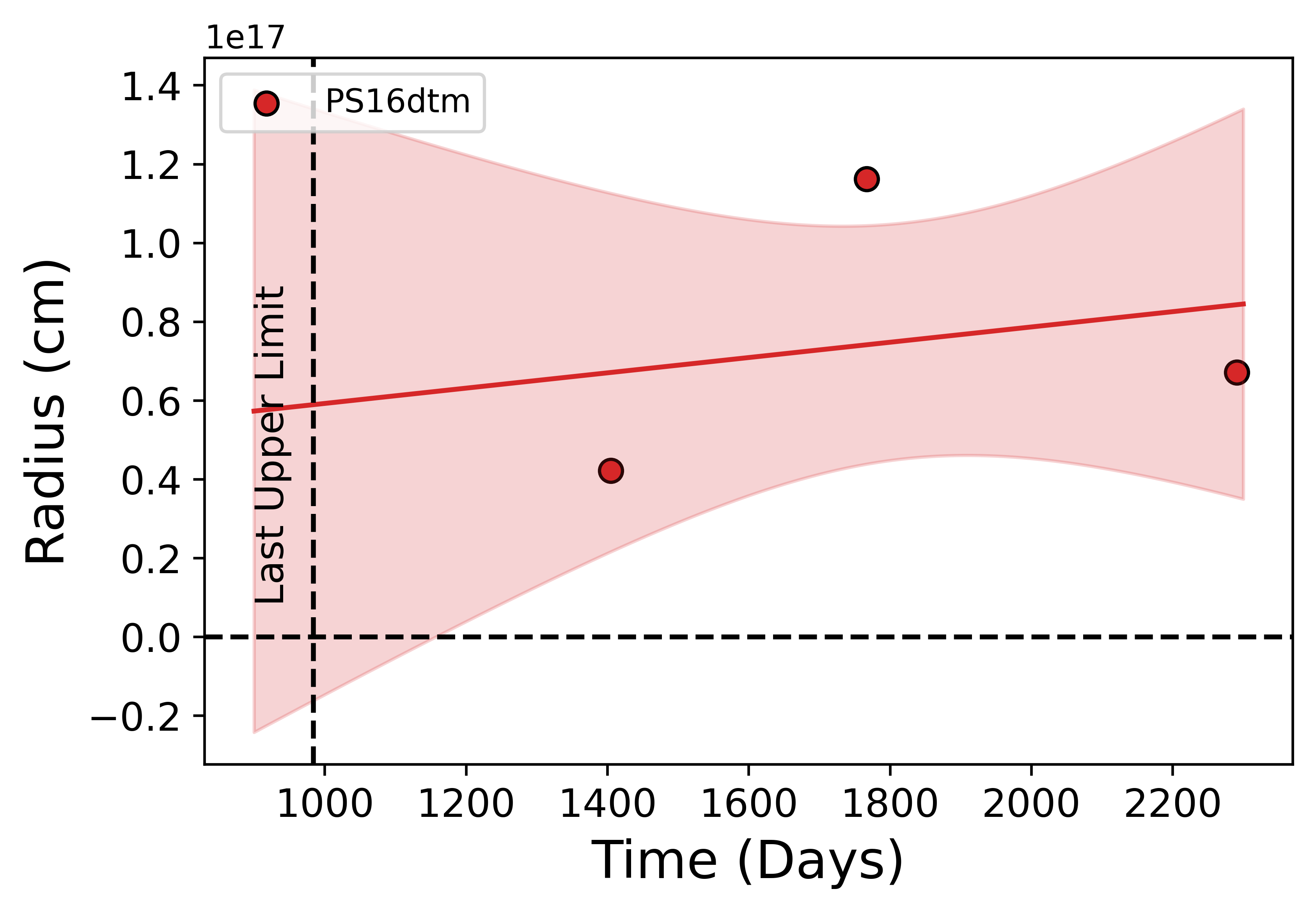}
    \includegraphics[width=.32\columnwidth]{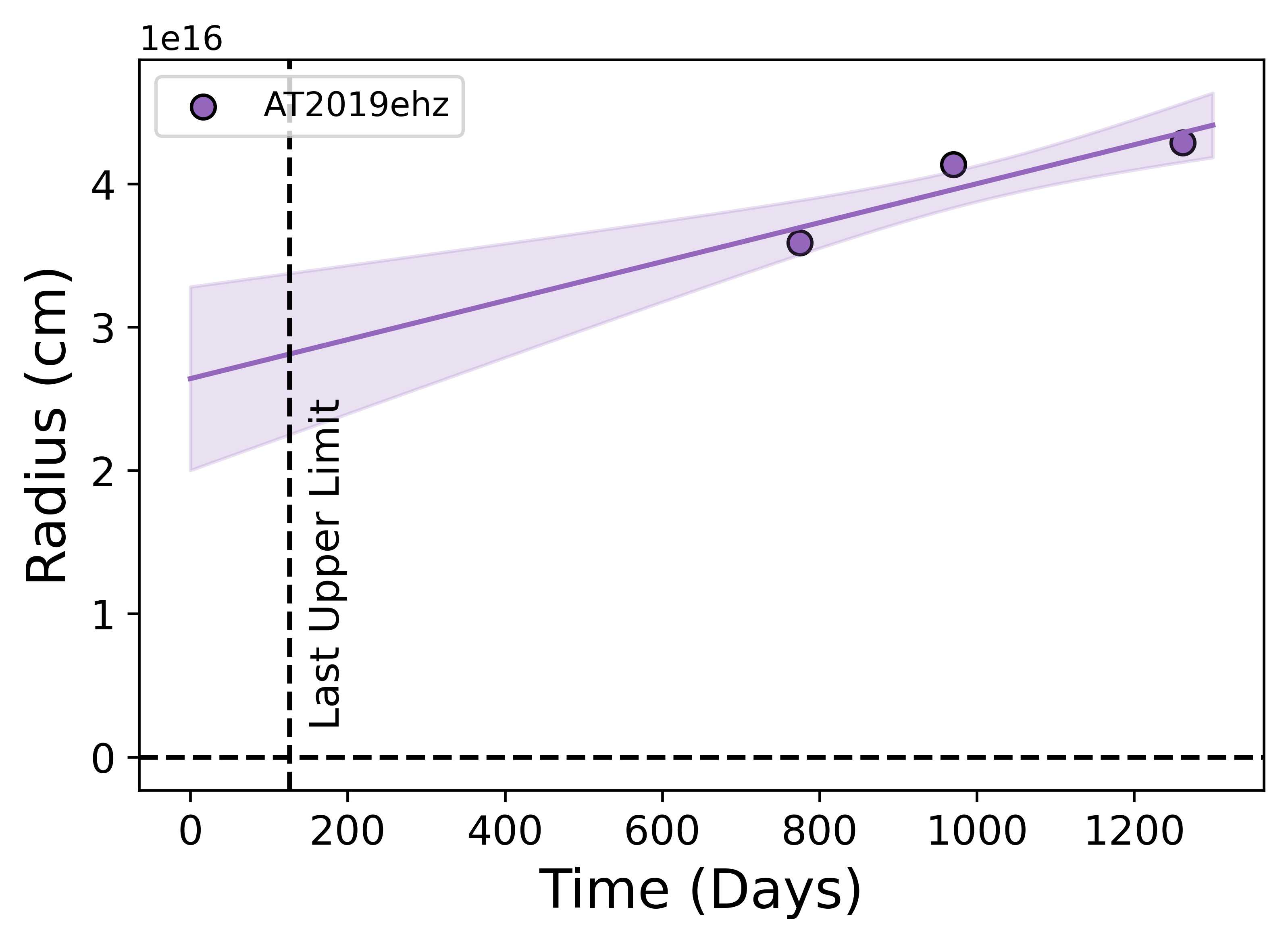}
    \label{fig:disruption}
    \caption{Equipartition radius as a function of time for each TDE with multiple SEDs.  The lines are linear fits to the radius evolution to determine the outflow launch time.  For reference we include the date of the last upper limit (vertical dashed lines).  The first data point for each TDE is inferred from a 6 GHz detection and the subsequent evolution of the SEDs (see \S\ref{sec:outflow-date}). For AT2019eve, we exclude the last SED point and show it as an open circle.}
\end{figure*}   

\begin{deluxetable*}{lllllr}
\label{tab:delayed}
\tablecolumns{6}
\tablecaption{Parameters With Delayed Outflows}
\tablehead{
	\colhead{TDE} &
	\colhead{Outflow Start} &
	\colhead{$\delta t_{\rm ori}$} &
	\colhead{$\delta t_{\rm rev}$} &
	\colhead{$\beta$} &
	\colhead{log($\nu_c$)} \\
	\colhead{} &
	\colhead{(d)} &
	\colhead{(d)} &
	\colhead{(d)} &
	\colhead{} &
	\colhead{(Hz)}
}
 \startdata
 \multicolumn{6}{c}{\emph{TDEs with Launch Times Estimated from Radius Evolution}} \\ 
ASASSN-14ae & $<2064\substack{+3 \\ -6}$ & 2696& 632 & >0.025 & >10.36 \\
ASASSN-14ae & $<2064\substack{+3 \\ -6}$ &3243& 1179 & >0.028 & >10.39 \\
AT2018hco & $790\substack{+70 \\ -110}$ &1191 &401& $0.039\substack{+0.016 \\ -0.015}$ & $10.42\substack{+0.04 \\ -0.04}$ \\
AT2018hco & $790\substack{+70 \\ -110}$ &1311 &521& $0.043\substack{+0.010 \\ -0.011}$ & $10.79\substack{+0.03 \\ -0.03}$ \\
AT2019eve & $545\substack{+30 \\ -30}$ &945 &400&  $0.139\substack{+0.012 \\ -0.014}$ & $11.55\substack{+0.11 \\ -0.13}$ \\
AT2019eve & $545\substack{+30 \\ -30}$ &1102 &557&  $0.153\substack{+0.014 \\ -0.013}$ & $11.82\substack{+0.11 \\ -0.11}$ \\
AT2019eve & $545\substack{+30 \\ -30}$ &1325 &780&  $0.094\substack{+0.016 \\ -0.016}$ & $11.43\substack{+0.11 \\ -0.11}$ \\
\hline
\multicolumn{6}{c}{\emph{TDEs with Individually Inferred Launch Times}} \\ 
iPTF16fnl & 1345 &1752 &407&  >0.002 & >9.45 \\
PS16dtm & 984 &1767& 783& >0.030 & >10.30 \\
PS16dtm & 984 &2291&1307& >0.023 & >10.45 \\
AT2018zr & 1218 &1743 &525&  >0.043& >11.09 \\
AT2019dsg & 561 &1171 &610&  $0.017\substack{+0.002 \\ -0.002}$ & $10.25\substack{+0.03 \\ -0.04}$ \\
AT2018dyb & 1028 &1615 &270&  $0.023\substack{+0.002 \\ -0.002}$ & $11.32\substack{+0.11 \\ -0.11}$ \\
AT2019ehz & 400 &945 &545&  $0.029\substack{+0.001 \\ -0.001}$ & $10.01\substack{+0.11 \\ -0.11}$ \\
AT2019ehz & 400 &1262 &862&  $0.020\substack{+0.002 \\ -0.002}$ & $10.23\substack{+0.11 \\ -0.11}$ \\
AT2019teq & 351 &1155 &804&  >0.080& >11.51 \\
 \enddata
 \tablecomments{The outflow start times are in days post optical discovery, where $\delta t_{\rm ori}$ are the original outflow dates using the date since optical detection, and $\delta t_{\rm rev}$ are the revised times of our SED observations relative to the outflow start times.
 }
 \end{deluxetable*}
 
In Figure~\ref{fig:disruption}, we show our fits to the radius data for each TDE (including for reference the date of the last measured upper limit).  In all cases, with the exception of AT2019ehz and PS16dtm, we find significant delays in the outflow launch time, of $\approx 545-2150$ days; see Table~\ref{tab:delayed}.  Equally important, the inferred time delays are consistent with with the available radio non-detections.  In the case of AT2019ehz, the data do not converge to a launch time because the decrease in peak flux and peak radius corresponds with a very small increase in radius.  It is possible, given that it was brightest when first detected, that the outflow had already decelerated, which is why the radius is not zero at the launch date in our analysis.  Deceleration of the outflow could also potentially explain the lack of convergence in our data for PS16dtm.  For ASASSN-14ae, we note since our radii used in this calculation are upper limits, the inferred launch date is also a limit, and could be even earlier.
 
In the case of TDEs with only a single SED, and for AT2019ehz and PS16dtm where our multi-epoch radius analysis did not converge, we instead make individual estimates of the launch date using the observed light curve behavior and prior upper limits.  For AT2019dsg, which exhibits a distinct late-time component, we use an outflow launch time of 561 days corresponding to the dimmest light curve point before the re-brightening.  For AT2018zr, with a constraining upper limit at 1218 days, we assume the outflow launched at that time.  For PS16dtm, we adopt the last constraining upper limit in radio observations at 984 days. For AT2019teq, with a constraining upper limit at 351 days, we assume the outflow launched at that time.  For AT2018dyb, which has brightened by an order of magnitude between our two observations, we assume the outflow launch time corresponds to our initial detection at 1028 days.  Finally, for AT2019ehz, which has a data gap of $\approx 730$ days between the final upper limit and the first detection, we assume that the light curve rose at roughly the same rate as its observed decline, giving an estimated launch time of $\sim 400$ days.  We note that while we are currently limited for these TDEs by the cadence of observations planned future observations of the TDEs with only a single SED will allow us to better refine the outflow launch times, and potentially increase our precision with estimates for TDEs in Figure~\ref{fig:disruption}.

With the outflow delay estimates we recalcualte the values of $\beta$ and $\nu_c$, which are strongly dependent on the chosen launch time (Table~\ref{tab:delayed}).  We find that this increases the inferred velocities by a factor of $\approx 2-5$, to $\beta\approx 0.02-0.15$.

Finally, we note that for all but one TDE, the revised $\nu_c$ is more in line with the data, and include these values here as a consistency check. Specifically, in most cases if we calculate $\nu_c$ from the time of optical discovery (Table~\ref{tab:equi}), we would expect a cooling break to be present in the range of frequencies covered by our SEDs.  However, no such break is discernible, implying the cooling break is higher than calculated.  The exception to this is AT2019ehz, for which an estimated launch date was not possible via a fit to its radius evolution. This discrepancy in $\nu_c$, resolves itself once we use shorter $\delta t$ in its calculation, however, which implies a delayed outflow.  However, other explanations are also possible due to the parameters required to calculate $\nu_c$, such as a deviation from equipartition.  We note that even if a deviation from equipartition is present, it minimally affects the radius (see Equation~\ref{eq:rad}), and thus would not significantly change our estimated outflow times.

\section{Discussion}
\label{sec:discussion}

\subsection{Outflow and Environment Properties of TDEs with Late Radio Emission}

\begin{figure}[t!]
    \centering \includegraphics[width=.8\columnwidth]{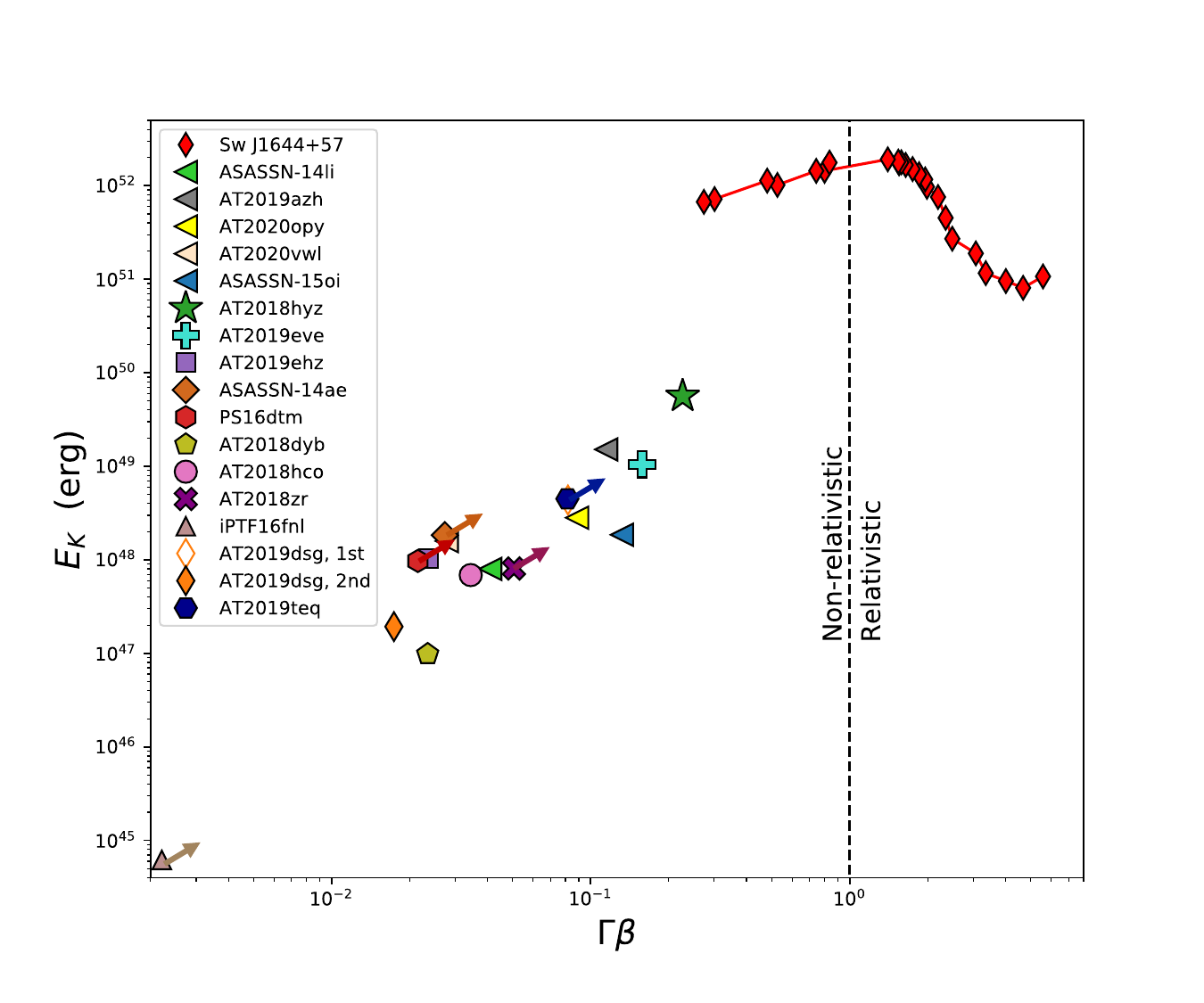}
    \label{fig:ek-gamma}
    \caption{Kinetic energy versus velocity for the new TDEs presented in this work (and AT2018hyz; \citealt{Cendes2022}). We use the inferred launch times in Table~\ref{tab:delayed}, and where our observations resulted in an upper limit we include include an arrow in the direction of allowable energy/velocity phase space.  For comparison we include optically-discovered TDEs with early radio emission (sideways triangles; \citealt{Cendes2021b,Stein2021,Alexander2016,Goodwin2022,Goodwin2023B,Goodwin2023}) and Sw1644+57 (red diamonds; \citealt{Zauderer2011,Cendes2021}).  In the case of AT2019azh and AT2020opy \citep{Goodwin2022,Goodwin2023B}, we show the peak energy and velocity adjusted to $\epsilon_B = 0.1$.  For ASASSN-15oi \citep{Horesh2021}, we use the observation with the highest peak frequency and peak luminosity (182 days) with $\epsilon_B = 0.1$ and $p=2.39$, which best fit the observed SED, and infer the velocity by subtracting 90 days \citep[the last date of non-detection; see][]{Horesh2021}.  We also include the peak energy and velocity in the first peak for AT2019dsg for reference (open orange diamond; \citealt{Cendes2021b,Stein2021}.}
\end{figure}

Our equipartition analysis for 8 TDEs with delayed radio emission and 2 TDEs with a second flare allows us to examine their energy and velocity relation, and to compare their properties to TDEs with early radio emission. In Figure~\ref{fig:ek-gamma} we plot the kinetic energy ($E_K$) and velocity ($\Gamma\beta$) for all known optical TDEs with radio detections for which a similar analysis has been carried out, using the highest energy inferred in those sources from Table \ref{tab:equi} and the literature\citep{Cendes2021b,Stein2021,Alexander2016,Goodwin2022,Zauderer2011,Cendes2021,Goodwin2023B,Cendes2022}.  We also include the jetted, non-optical TDE Sw1644+57 for reference \citep{Eftekhari2018,Cendes2021}.  For TDEs where we obtained lower limits for these values, we include an arrow in the direction of the allowable phase space.  For TDEs with late emission, with the exception of iPTF16fnl (which had a lower limit at $E_K > 10^{45}$ erg and $\Gamma\beta > 0.002$), we find they span $E_K\approx 10^{47}-10^{49}$ erg, and $\Gamma\beta\approx 0.01-0.1$.  Overall, we find the majority of the TDE population in our sample show velocity and energy values similar to that seen in a broader population of TDEs with non-relativistic outflows that have emission at early times \citep[$\sim10^{47}-10^{49}$; i.e., ][]{Cendes2021b,Alexander2016}.  

\begin{figure}
    \centering\includegraphics[width=.7\columnwidth]{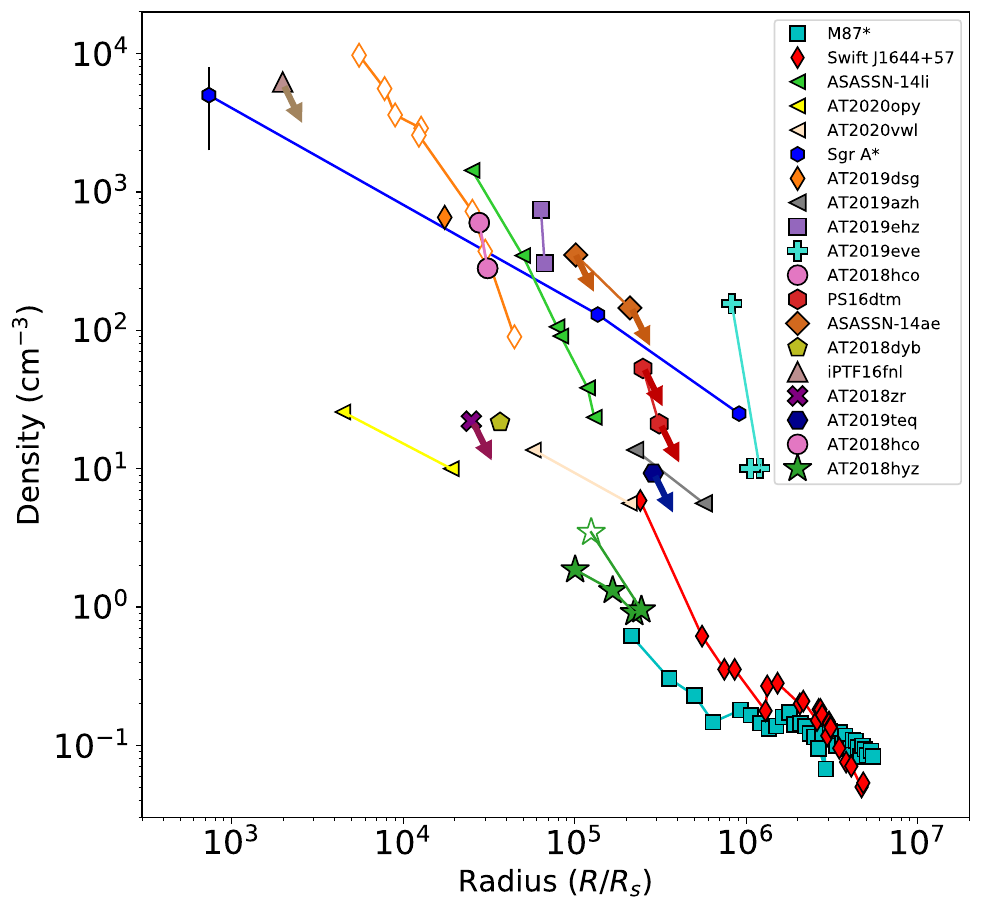}
    \label{fig:density}
    \caption{The circumnuclear density profiles inferred for TDEs presented in this work (and AT2018hyz; \citealt{Cendes2021b}), normalized to the Schwarzschild radius of each SMBH.  We also include for comparison the density profiles for optically-discovered TDEs with early radio emission (sideway triangles; \citealt{Alexander2016,Cendes2021b,Horesh2021,Goodwin2022,Goodwin2023,Goodwin2023B}), Sw1644+57 \citep{p1,p2,Eftekhari2018,Cendes2021}, and the Milky Way \citep{Baganoff2003,Gillessen2019} and M87 \citep{Russell2015}.  For TDEs where our density value is an upper limit, we include an arrow in the direction of the phase space of allowable values.
}
\end{figure}

In Figure~\ref{fig:density} we show the inferred ambient density for each TDE as a function of radius (scaled by the Schwarzschild radius) for TDEs in our sample and previously studied radio-emitting TDEs.  Here we use the mass for each BH inferred from optical data \citep{Nicholl2019,Yao2022,Goodwin2022,Goodwin2023B,Leloudas2019,Blagorodnova2017,Hammerstein2023}.  In the cases of TDEs where we have upper limits on density and lower limit on radius, we include an arrow in the direction of the allowable phase space.  We find that the densities probed are consistent with the densities and circumnuclear density profiles of previous TDEs, and are more dense than the low density environment of M87$^*$.  Crucially, we do not infer an unusually high density for any TDEs in our sample, which might be expected if the radio emission was delayed due to rapid shift from low to high density.

As many of our light curves are still rising, this implies that the blast wave is still in the free expansion phase (with $v \approx$ constant).  In this case, the mass of the swept-up material, $M_{swept}$, would be less than the mass of the ejecta, $M_{ej}$.  Using the inferred ambient densities, kinetic energies, and velocities inferred from the equipartition analysis we verify that $M_{swept} \lesssim M_{ej}$ for all TDEs except for AT2019eve, where $M_{swept} \gtrsim M_{ej}$ in the first observation at 945 days and $M_{swept} \approx M_{ej}$ in our second observation at 1102 days.  Given the steady fading light curve of AT2019eve during this time period, we conclude AT2019eve is no longer in free expansion.

We can consider the case of the X-ray TDE IGRJ12580+0134, which had a radio-only flare observed $\sim1600$ days after peak \citep{Perlman2022}.  \citet{Perlman2022} interpret this flare is most likely due to a jet outflow encountering a circumnuclear (CNM) cloud at $\approx2.9-3\times10^{18}$ cm from a $\sim10^5 M_{\odot}$ black hole \citep{Irwin2015}, where the density increases from a background of $5.2$ cm$^{-3}$ to $10.3$ cm$^{-3}$.  These densities would be lower than what we sample in our TDE population and at larger radii than we probe in our work ($R/R_s > 10^{7}$, and the outflow velocity of $\beta \approx 0.17$ is higher than what we infer for all our TDEs.  However, as \citep{Perlman2022} notes theirs is a simplified model with several degeneracies, and other values may fit the data.

\subsection{Off-Axis Jets}
\label{sec:jet}

We explore the possibility that the late radio emission is due to relativistic jets with an off-axis viewing orientation.  First, we note that if the origin of delayed radio emission was off-axis relativistic jets, then the inevitable conclusion based on our detection statistics (\S\ref{sec:rate}) is that about half of all optically-selected TDEs harbor off-axis jets.  This conclusion, which would indicate a rate of $\sim {\rm few}\times 10^2$ Gpc$^{-3}$ yr$^{-1}$ (using the optical TDE rate from \citealt{vanVelzen2018}) is in strong conflict with the rate of on-axis relativistic jets determined from $\gamma$-rays and optical detections \citep{Andreoni2022,Brown2015} of $\sim 0.01-0.02$ Gpc$^{-3}$ yr$^{-1}$ and a beaming correction factor of $\sim 10^2$.  

Further, we can consider two possible off-axis jet scenarios: (i) an initially off-axis relativistic jet that has decelerated to a non-relativistic velocity and spread to roughly spherical geometry, which is mainly relevant for TDEs in our sample in which we have observed the light curves peak and decline; and (ii) an off-axis jet that remains collimated and oriented off-axis, which is potentially relevant for TDEs in which the radio emission is still rising (e.g., \citep{Matsumoto2023}).  

In the first scenario, the time at which the radio emission peaks is given by the deceleration time (e.g., \citealt{Nakar2011}):
\begin{equation}
t_{\rm dec} \approx 30\, {\rm d}\,\, E_{\rm eq,49}^{1/3} n_{\rm ext}^{-1/3} \beta_0^{-5/3}\\
\label{eq:tdec}
\end{equation}
where $\beta_0$ is the initial velocity, which for an off-axis relativistic jet is $\beta_0=1$. Using the results of our equipartition analysis (\S\ref{sec:equi}), we find $t_{\rm dec}\lesssim 100$ days for all TDEs in our sample, which is substantially smaller than the observed peak timescales of $\gtrsim 700$ days. This agrees with the much lower luminosities compared to Sw1644+57, which became non-relativistic at a comparable timescale of $\approx 700$ d \citep{Eftekhari2018}. In particular, several of the TDEs in our sample (i.e., PS16dtm, AT2018hco, AT2019ehz, and AT2019eve) peaked at $\sim 700-1700$ days with luminosities of $\sim 10^{38}-10^{39}$ erg s$^{-1}$, about two to three orders of magnitude below Sw1644+57.  These TDEs consistently have values of $\beta\approx 0.01-0.07$ and have already peaked, so following the formalism of  \citet{Matsumoto2023} and \citet{Beniamini2023} we conclude that they cannot be off-axis jets since they do not cross the threshold of $\beta\approx 0.44$ required for a decelerating off-axis jet.

The second scenario involves the possibility of a decelerating off-axis jet that remains collimated with the emitting area increasing over time, as suggested for the delayed rapid rise in AT2018hyz by \citet{Matsumoto2023} and \citet{Sfaradi2023}. In updated work, \citet{Beniamini2023} proposes a threshold of $\beta\approx 0.44$ below which two solutions exist: a non-relativistic spherical outflow, and an off-axis relativistic jet.  Critically, in the latter scenario, as the jet decelerates and the emission area grows, the outflow will eventually exceed the threshold value. For the TDEs in our sample with rising light curves (i.e., ASASSN-14ae, AT2018dyb, AT2018zr) we find values for $\beta\approx 0.01\ll 0.44$.  Since these sources still exhibit rising emission, continued monitoring will establish whether they ever cross the threshold value of $\beta$.  However, this is unlikely given their current evolution.  For example, in the case of ASASSN-14ae, which is rising rapidly after a long delay of $\gtrsim 2300$ days, we find that with its current evolution ($F_{\nu,p}\propto t^{4.4}$, $\nu_p\approx {\rm const}$) it is expected to follow $\beta\propto t^{1.1}$, and hence will cross the threshold of 0.23 (0.44) in $\approx 17$ ($\approx 31$) years given the current value of $\beta\approx 0.01$; at that point it will reach a peak luminosity of $\sim {\rm few}\times 10^{43}$ erg s$^{-1}$, much in excess of the peak lumiosity of Sw1644+57.  Similar conclusions are reached for AT2018dyb and AT2018zr, which begin rising at $\approx 10^3$ days.

We next note the case of AT2019dsg, which exhibits early radio emission (i.e., peaking at $\sim 200-400$ days) and then rapidly rises again a second time.  We could envision a scenario in which this TDE produced both a spherical non-relativistic outflow that dominates the early radio emission, and an off-axis relativistic jet.  The inferred velocities for this TDE is low, $\beta\approx 0.003-0.009$, and hence they are again unlikely to eventually cross the threshold of $\beta\approx 0.44$ expected for an off-axis jet.  As for the other TDEs with rising light curves, continued monitoring of AT2019dsg will elucidate the origin of its emission.

Thus, we conclude based on the measured radio properties (timescales and luminosities), the inferred physical properties (velocities), and the rate of TDEs with late radio emission that off-axis jets are an unlikely explanation for this population.

\subsection{Origin of Delayed Outflows}

There are at least two broad possibilities for the origin of the delayed mildly relativistic outflow, both of which connect to its assumed origin in a fast outflow from the innermost regions of the black hole accretion flow.  We note explicitly that it is possible that late-time radio emission is of diverse origins, given the range of behavior in terms of parameters such as the timescales, luminosities, rise rates, and other features.  Additional observations of radio TDEs at late times will allow us to distinguish between scenarios.

One possibility is that the delayed radio emission is due to the timescales for debris circularization and viscous accretion \citep{Hayasaki2021}.  In this scenario, the first stream-stream collisions produce the optical/UV flare, creating a debris-circularized ring.  The ring then evolves viscous-diffusively, and reaches the innermost stable circular orbit (ISCO) on a timescale of months to years after the initial flare, consistent with the timescale we infer for our TDEs.  However, with a disk wind velocity of $\sim 0.4c$, this is inconsistent with the velocities we infer for our TDEs, although may be possible to achieve if the outflow mass was very low or the radius for the ISCO was exceptionally large.

Alternately, it is also possible that the formation of a jet or fast wind from the inner accretion flow is delayed because the SMBH accretion rate does not peak on the same timescale as the mass fall-back rate.  Because the debris from the TDE is weakly bound to the SMBH, its natural configuration is a large quasi-spherical envelope (e.g., \citealt{Loeb1997,Coughlin2014}), which must cool and radially contract to form an accretion disk (e.g., \citealt{Metzger2022}).  If accretion onto the SMBH supplies energy to the envelope (with an efficiency $\eta$), full contraction of the envelope can be delayed for a time, up to $\sim$700 days (e.g., \citealt{Loeb1997}), in agreement with the time scales of the outflows measured in this work.  Thus, we see that disk formation can be delayed for hundreds to thousands of days, providing an alternative explanation for late-onset radio-generating outflows from the SMBH \citep{Metzger2022}.  This model also appears to provide a good description of the multi-wavelength properties of ASASSN-15oi (Hajela et al. in prep).  As another alternative possibility, the escape of a jet from the vicinity of the black hole may be delayed until processes within the accretion flow align the angular momentum axis of the accretion disk and the jet to the black hole spin axis \citep{Teboul2023}.

We note explicitly that it is possible that late-time radio emission is of diverse origins, given the range of behavior in terms of parameters such as the timescales, luminosities, rise rates, and other features.  Additional observations of radio TDEs at late times will allow us to distinguish between scenarios.

\section{Conclusions}
\label{sec:conclusions}

We presented radio observations for 23 optically-discovered TDEs on timescales of $\approx 500-3200$ days post-discovery.  We detected radio emission from 17 of these TDEs, of which 6 had an ambiguous or host/AGN origin.  Of the 11 TDEs with transient radio emission, 9 TDEs were detected for the first time despite a lack of radio emission at earlier times, and 1 TDE (AT2019dsg)  was detected to significantly brighten at late time relative to its declining radio emission at earlier times; this late-time component is similar to what has been found in ASASSN-15oi and AT2020vwl \citep{Horesh2021,Goodwin2023,Goodwin2023C}. Based on this large sample, our key results are:

\begin{itemize}

\item We find $\approx40\%$ of TDEs in this work show late-time emission, meaning a substantial fraction of all optically-selected TDEs exhibit radio emission that rises on timescales of hundreds of days.

\item The range of luminosities for this sample is $\sim 10^{37}-10^{39}$ erg s$^{-1}$, though in some cases this is a lower limit because emission is still rising.

\item We find radio emission in this population peaks at time scales $700-3200$d, though note some are lower limits as emission is still rising.

\item Multi-frequency SEDs reveal a range of peak frequencies of $\lesssim 1$ GHz to $\approx 5$ GHz. Using the SED information we determine the outflow physical properties assuming equipartition. We find $R_{\rm eq}\approx 3\times10^{16}- 2\times10^{17}$ cm and $E_K\approx 8\times10^{46}- 1\times10^{49}$ erg.

\item Using the radius evolution when available, and the light curve behavior otherwise, we infer outflow launch timescales of $\approx 500-2000$ days. This then leads to inferred velocities of $\beta\approx 0.02-0.15$.

\item From the equipartition analysis we also infer circumnuclear densities of $\approx 10^{1}-10^{4}$ cm$^{-3}$.  These densities are comparable to those inferred for TDEs with early radio emission.

\item We rule out off-axis relativistic jets as an explanation for the bulk of the TDEs with late radio emission, and conclude delayed outflows are a more likely explanation.  If this delay is due to delayed disk formation, then the relative formation timescale inferred from our data is $\sim 700$ days.
 
\end{itemize}
   
Our study highlights that persistent radio monitoring of TDEs starting after discovery and lasting for several years is critical for determining the timing of ourflow formation, and its subsequent evolution; this includes TDEs that exhibit radio emission at early time, but which may re-brighten subsequently.  Additionally, multi-frequency observations, extending below $\sim 1$ GHz, are crucial for determining the outflow properties (energy, velocity) and the ambient density.  As demonstrated here, these observations are within reach for TDEs at modest distances with existing facilities (VLA, ATCA, MeerKAT).  In the future, much larger samples of TDEs (typically at larger distances) from the Vera C.~Rubin Observatory Legacy Survey of Space and Time (LSST) will be effectively studied with the next-generation Square Kilometer Array and the next-generation VLA.

\begin{acknowledgments}
We thank Tatsuya Matsumoto and Nick Stone for their helpful discussions on TDEs.  The Berger Time-Domain Group at Harvard is supported by NSF and NASA grants. The National Radio Astronomy Observatory is a facility of the National Science Foundation operated under cooperative agreement by Associated Universities, Inc.  The MeerKAT telescope is operated by the South African Radio Astronomy Observatory, which is a facility of the National Research Foundation, an agency of the Department of Science and Innovation.  The Australia Telescope Compact Array is part of the Australia Telescope National Facility (https://ror.org/05qajvd42) which is funded by the Australian Government for operation as a National Facility managed by CSIRO.  This research has made use of the CIRADA cutout service at URL cutouts.cirada.ca, operated by the Canadian Initiative for Radio Astronomy Data Analysis (CIRADA). CIRADA is funded by a grant from the Canada Foundation for Innovation 2017 Innovation Fund (Project 35999), as well as by the Provinces of Ontario, British Columbia, Alberta, Manitoba and Quebec, in collaboration with the National Research Council of Canada, the US National Radio Astronomy Observatory and Australia’s Commonwealth Scientific and Industrial Research Organisation.
\end{acknowledgments}

\appendix

\startlongtable
\begin{deluxetable}{llllllr}
\tablecolumns{7}
\tablecaption{Radio Observations of TDEs}
\tablehead{
Object&
Telescope&
Project Code/ &
Date of Observation&
$\delta t$ &  
Frequency  & 
Flux Density \\
&
& Source
& 
& 
(d)   & 
(GHz) & 
(mJy) 
}
\startdata
\hline
iPTF16fnl & VLA & 20A-492 & 2020 May 24 & 1365 & 6 & 0.045 $\pm$ 0.001\\
& --- & 21A-303& 2021 Jun 15 & 1752 & 3 & $<$0.514\\
& ---& --- & --- & --- & 5 & 0.028 $\pm$ 0.008\\
& ---& ---& --- & --- & 7 & 0.031 $\pm$ 0.007\\
& ---& ---& --- & --- & 9 & 0.024 $\pm$ 0.006\\
& ---& ---& --- & --- & 11 & 0.023 $\pm$ 0.008\\
\hline
AT2019dsg & VLA & 21A-303 & 2021 Jun 12 & 796 & 1.25 & $<$0.514\\
& --- & ---& --- & --- & 1.75 & 0.114 $\pm$ 0.033\\
& --- & ---& --- & --- & 2.5 & 0.235 $\pm$ 0.026\\
& --- & ---& --- & --- & 3.5 & 0.201 $\pm$ 0.014\\
& ---& --- & --- & --- & 5 & 0.209 $\pm$ 0.014\\
& ---& ---& --- & --- & 7 & 0.132 $\pm$ 0.013\\
& MeerKAT & SCI-20220822-YC-01 & 2023 Jan 15 & 1378 & 0.88 & 0.284 $\pm$ 0.041\\
& ---& --- & --- & --- & 1.36 & 0.384 $\pm$ 0.026\\
& VLA& VLASS 3 & 2023 Jan 31 & 1404 & 3 & $<$0.386\\
& MeerKAT & SCI-20220822-MB-03 & 2023 Mar 5 & 1437 & 1.36 & 0.228 $\pm$ 0.016\\
\hline
ASASSN-14ae & VLA& 15B-247 & 2016 Mar 12 & 778 & 5 & $<$0.033\\
& ---& VLASS 1 & 2019 Nov 19 & 2122 & 3 & $<$0.420\\
& ---& 20A-492 & 2020 May 25 & 2313 & 6 & 0.090 $\pm$ 0.015\\
& ---& 21A-303& 2021 Jun 12 & 2696 & 3 & 0.334 $\pm$ 0.038\\
& ---& --- & --- & --- & 5 & 0.222 $\pm$ 0.018\\
& ---& ---& --- & --- & 7 & 0.169 $\pm$ 0.007\\
& ---& ---& --- & --- & 9 & 0.155 $\pm$ 0.010\\
& ---& ---& --- & --- & 11 & 0.150 $\pm$ 0.001\\
& ---& VLASS 2 & 2021 Dec 4 & 2608 & 3 & $<$0.47\\
& ---& 22B-205 & 2022 Dec 11 & 3243 & 1.25 & $<$3.12\\
& ---& --- & --- & --- & 1.75 & 0.720 $\pm$ 0.041\\
& ---& ---& --- & --- & 2.5 & 0.680 $\pm$ 0.029\\
& ---& ---& --- & --- & 3.5 & 0.604 $\pm$ 0.017\\
& ---& ---& --- & --- & 5 & 0.418 $\pm$ 0.080\\
& ---& --- & --- & --- & 7 & 0.446 $\pm$ 0.011\\
& ---& ---& --- & --- & 9 & 0.382 $\pm$ 0.020\\
& ---& ---& --- & --- & 11 & 0.361 $\pm$ 0.025\\
& ---& ---& --- & --- & 13.5 & 0.347 $\pm$ 0.015\\
& ---& ---& --- & --- & 16.5 & 0.337 $\pm$ 0.022\\
& ---& ---& --- & --- & 20 & 0.303 $\pm$ 0.015\\
& ---& ---& --- & --- & 24 & 0.343 $\pm$ 0.014\\
\hline
%PS16dtm & VLA & 16B-398 & 2016 Sept 22 & 54 & 15.5 & $<$0.069\\
% & ---& --- & 2016 Dec 21 & 144 & 6 & $<$0.075\\
PS16dtm & VLA & 16B-318 & 2017 Jun 4 & 326 & 6 & $<$0.0303\\
 & --- & --- & 2017 Jun 4 & 326 & 21.7 & $<$0.0663\\
 & ---& --- & 2017 Aug 22 & 372 & 6 & $<$0.026\\
  & ---& --- & 2017 Aug 22 & 372 & 21.7 & $<$0.075\\
 & ASKAP&  & 2019 Jun 16 & 984 & 0.887 & 1.002 $\pm$ 0.015\\
 & VLA& 20A-492 & 2020 Jun 6 & 1405 & 6 & 0.210 $\pm$ 0.008\\
 & ---& VLASS 2 & 2020 Sept 4 & 1485 & 3 & $<$0.845\\
& ---& 21A-303& 2021 Jun 14 & 1767 & 3 & 0.361 $\pm$ 0.076\\
& ---& --- & --- & --- & 5 & 0.285 $\pm$ 0.013\\
& ---& ---& --- & --- & 7 & 0.260 $\pm$ 0.009\\
& ---& ---& --- & --- & 9 & 0.256 $\pm$ 0.010\\
& ---& ---& --- & --- & 11 & 0.228 $\pm$ 0.009\\
& ---& 22B-205 & 2022 Nov 23 & 2291 & 1.25 & $<$0.767\\
& ---& --- & --- & --- & 1.75 & 0.193 $\pm$ 0.047\\
& ---& ---& --- & --- & 3.0 & 0.210 $\pm$ 0.058\\
& ---& ---& --- & --- & 6 & 0.163 $\pm$ 0.012\\
& ---& ---& --- & --- & 10 & 0.089 $\pm$ 0.015\\
& ---& ---& --- & --- & 15 & 0.086 $\pm$ 0.008\\
& ---& ---& --- & --- & 20 & 0.053 $\pm$ 0.012\\
& ---& ---& --- & --- & 24 & 0.039 $\pm$ 0.011\\
\hline
AT2018zr & VLA& 20A-492 & 2020 Aug 23 & 929 & 6 & $<$0.014\\
& VLA& 21A-303 & 2021 Jun 18 & 1218 & 6 & $<$0.053\\
& --- & 22B-205 & 2022 Oct 16 & 1713 & 6 & 0.147 $\pm$ 0.011\\
& ---& --- & 2022 Dec 8 & 1743 & 1.25 & $<$0.749\\
& ---& --- & --- & --- & 1.75 & 0.218 $\pm$ 0.041\\
& ---& ---& --- & --- & 2.5 & 0.209 $\pm$ 0.032\\
& ---& ---& --- & --- & 3.5 & 0.180 $\pm$ 0.001\\
& ---& ---& --- & --- & 5 & 0.155 $\pm$ 0.018\\
& ---& --- & --- & --- & 7 & 0.109 $\pm$ 0.015\\
& ---& ---& --- & --- & 9 & 0.088 $\pm$ 0.016\\
& ---& ---& --- & --- & 11 & 0.085 $\pm$ 0.012\\
\hline
%AT2020neh & VLA& 20A-372 & 2020 Jun 30 & 12 & 15& $<$0.018\\
%& ---& --- & 2020 Dec 31 & 196 & 6 & $<$0.016\\
AT2020neh & VLA & 22B-205 & 2022 Nov 9 & 874 & 6 & 0.026 $\pm$ 0.006\\
& ---& --- & 2022 Dec 10 & 905 & 6 & 0.053 $\pm$ 0.012\\
\hline
AT2018dyb & MeerKAT & SCI-20210212-YC-01 & 2021 May 4 & 1028 & 1.36 & 0.158 $\pm$ 0.045\\
& MeerKAT & SCI-20220822-YC-01 & 2022 Dec 11 & 1615 & 1.36 & 1.031 $\pm$ 0.068\\
& ATCA & C3325 & 2023 Jan 22 & 1657 & 2.1 & 1.15 $\pm$ 0.07\\
& ---& ---& --- & --- & 5.5 & 0.50 $\pm$ 0.04\\
& ---& --- & --- & --- & 9 & 0.27 $\pm$ 0.03\\
\hline
AT2018hco & VLA & 18A-373 & 2018 Dec 5 & 62 & 6 & $<$0.0165\\
 & --- & VLASS 1&2019 May 5 & 213 & 3 & $<$0.3\\
 & --- & 21A-303 & 2021 June 12 & 982 & 6 & 0.343 $\pm$ 0.015\\
 & --- & VLASS 2& 2021 Oct 14 & 1106 & 3 & 0.436 $\pm$ 0.142\\
 & ---& 21B-360 & 2022 Jan 7 & 1191 & 1.5 & $<$0.329\\
& ---& ---& --- & --- & 2.5 & 0.166 $\pm$ 0.030\\
& ---& ---& --- & --- & 3.5 & 0.252 $\pm$ 0.011\\
& ---& ---& --- & --- & 5 & 0.265 $\pm$ 0.014\\
& ---& --- & --- & --- & 7 & 0.227 $\pm$ 0.014\\
& ---& ---& --- & --- & 9 & 0.161 $\pm$ 0.010\\
& ---& ---& --- & --- & 11 & 0.371 $\pm$ 0.015\\
& ATCA & ---& 2022 Apr 30 & 1311 & 2.1 & $<0.33$\\
& ---& ---& --- & --- & 5.5 & 0.200 $\pm$ 0.019\\
 & ---& --- & --- & --- & 9 & 0.111 $\pm$ 0.019\\
& ---& ---& --- & --- & 17 & $<0.084$\\
& ---& ---& --- & --- & 19 & $<$0.141\\
& MeerKAT & SCI-20210212-YC-01 & 2022 May 7 & 1318 & 0.88 & $<0.589$\\
& ---& ---& --- & --- & 1.36 & 0.171 $\pm$ 0.025\\
\hline 
AT2019ehz & VLA &VLASS 1& 2017 Dec 1 & -515 & 3 & $<$0.39\\
& --- & 19A-395 & 2019 May 21 & 23 & 9 & $<$0.06\\
& --- & 19A-395 &2019 June 14 & 47 & 9 & $<$0.026\\
& --- &VLASS 2 & 2019 Sept 1 & 126 & 3 & $<$0.3\\
& ---& 21A-303 & 2021 June 11 & 775 & 6 & 1.071 $\pm$ 0.032\\
& ---& 21B-360 & 2021 Dec 23 & 970 & 1.5 & $<$0.237\\
& ---& ---& --- & --- & 2.5 & 0.156 $\pm$ 0.070\\
& ---& ---& --- & --- & 3.5 & 0.396 $\pm$ 0.046\\
& ---& ---& --- & --- & 5 & 0.733 $\pm$ 0.041\\
& ---& --- & --- & --- & 7 & 0.485 $\pm$ 0.059\\
& ---& ---& --- & --- & 9 & 0.401 $\pm$ 0.014\\
& ---& ---& --- & --- & 11 & 0.338 $\pm$ 0.015\\
& ---& ---& --- & --- & 14 & 0.322 $\pm$ 0.015\\
& ---& ---& --- & --- & 17 & 0.200 $\pm$ 0.012\\
& ---& ---& --- & --- & 20 & 0.092 $\pm$ 0.016\\
& ---& ---& --- & --- & 24 & 0.077 $\pm$ 0.014\\
& ---& 22B-205 & 2022 Oct 7 & 1262 & 1.5 & $<$0.335\\
& ---& ---& --- & --- & 2.5 & 0.245 $\pm$ 0.026\\
& ---& ---& --- & --- & 3.5 & 0.309 $\pm$ 0.010\\
& ---& ---& --- & --- & 5 & 0.205 $\pm$ 0.032\\
& ---& --- & --- & --- & 7 & 0.235 $\pm$ 0.018\\
& ---& ---& --- & --- & 9 & 0.184 $\pm$ 0.013\\
& ---& ---& --- & --- & 11 & 0.111 $\pm$ 0.015\\
\hline
AT2019eve & VLA& VLASS 1 & 2017 Oct 28 & -555 & 3 & $<$0.3\\
 & --- & VLASS 2 & 2020 Oct 11 & 526 & 3 & $<$.0.497\\
 & ---& 21A-303 & 2021 June 11 & 769 & 6 & 0.766 $\pm$ 0.009\\
 & ---& 22B-360 & 2021 Dec 4 & 945 & 1.25 & 0.852 $\pm$ 0.120\\
& ---& --- & --- & --- & 1.75 & 0.978 $\pm$ 0.053\\
& ---& ---& --- & --- & 2.5 & 0.949 $\pm$ 0.039\\
& ---& ---& --- & --- & 3.5 & 0.847 $\pm$ 0.020\\
& ---& ---& --- & --- & 5 & 0.657 $\pm$ 0.022\\
& ---& --- & --- & --- & 7 & 0.484 $\pm$ 0.024\\
& ---& ---& --- & --- & 9 & 0.385 $\pm$ 0.034\\
& ---& ---& --- & --- & 11 & 0.222 $\pm$ 0.023\\
& ---& ---& --- & --- & 13.5 & 0.161 $\pm$ 0.023\\
& ---& ---& --- & --- & 16.5 & 0.171 $\pm$ 0.015\\
& ---& ---& --- & --- & 20 & 0.182 $\pm$ 0.011\\
& ---& ---& --- & --- & 24 & 0.123 $\pm$ 0.012\\
& MeerKAT & DDT-20220414-YC-01 & 2022 May 5 & 1103 & 0.88 & 0.702 $\pm$ 0.040\\
& ---& ---& --- & --- & 1.36 & 1.053 $\pm$ 0.020\\
& ATCA & C3472& 2022 Apr 30 & 1092 & 2.1 & 0.770 $\pm$ 0.061\\
& ---& ---& --- & --- & 5.5 & 0.543 $\pm$ 0.025\\
 & ---& --- & --- & --- & 9 & 0.300 $\pm$ 0.013\\
& ---& ---& --- & --- & 17 & 0.122 $\pm$ 0.032\\
& ---& ---& --- & --- & 19 & $<$.243\\
 & ---& 22B-205 & 2022 Dec 19 & 1325 & 1.25 & 0.846 $\pm$ 0.175\\
& ---& --- & --- & --- & 1.75 & 0.735 $\pm$ 0.048\\
& ---& ---& --- & --- & 2.5 & 0.553 $\pm$ 0.028\\
& ---& ---& --- & --- & 3.5 & 0.567 $\pm$ 0.018\\
& ---& ---& --- & --- & 5 & 0.430 $\pm$ 0.020\\
& ---& --- & --- & --- & 7 & 0.392 $\pm$ 0.015\\
& ---& ---& --- & --- & 9 & 0.272 $\pm$ 0.019\\
& ---& ---& --- & --- & 11 & $<$6\\
& ---& ---& --- & --- & 13.5 & $<$4.2\\
& ---& ---& --- & --- & 16.5 & 0.205 $\pm$ 0.010\\
& ---& ---& --- & --- & 20 & 0.076 $\pm$ 0.020\\
& ---& ---& --- & --- & 24 & 0.104 $\pm$ 0.016\\
& MeerKAT & SCI-20220822-YC-01 & 2023 Jan 4 & 1343 & 0.88 & 0.702 $\pm$ 0.040\\
\hline
AT2019teq & VLA& VLASS 2 & 2020 Aug 13 & 351 & 3 & $<$0.329\\
& --- & 22B-205 & 2022 Oct 19 & 1096 & 6 & 0.238 $\pm$ 0.008\\
 & ---& --- & 2022 Dec 17 & 1155 & 1.25 & 0.492 $\pm$ 0.088\\
& ---& --- & --- & --- & 1.75 & 0.430 $\pm$ 0.054\\
& ---& ---& --- & --- & 2.5 & 0.283 $\pm$ 0.051\\
& ---& ---& --- & --- & 3.5 & 0.268 $\pm$ 0.017\\
& ---& ---& --- & --- & 5 & 0.159 $\pm$ 0.023\\
& ---& --- & --- & --- & 7 & 0.064 $\pm$ 0.015\\
& ---& ---& --- & --- & 9 & $<$0.069\\
& ---& ---& --- & --- & 11 & 0.046 $\pm$ 0.018\\
\hline
DES14C1kia & VLA & 14B-506 & 2015 Jan 17 & 68 & 6 & $<0.017$\\
 & --- & --- & 2015 Jan 17 & 68 & 21.7 & $<0.033$\\
 & --- & --- & 2015 Mar 12 & 122 & 6 & $<0.015$\\
 & --- & --- & 2015 Mar 12 & 122 & 21.7 & $<0.043$\\
 & VLA & 20A-492 & 2020 May 28 & 2026 & 6 & $<0.013$\\
\hline
iPTF15af & VLA& 14A-483 & 2015 Jan 31 & 17 & 6.1 & $<$0.084\\
 & ---& 20A-492 & 2020 Jun 6 & 1970 & 6 & $<$0.017\\
 &--- & 22B-205 & 2022 Oct 25 & 2841 & 6 & $<$0.008\\
\hline
iPTF16axa& VLA& 21A-303 & 2021 Jun 17 & 1846 & 6 & $<$0.011\\
 & --- & 22B-205 & 2022 Oct 25 & 2841 & 6 & $<$0.017\\
\hline
AT2017eqx & VLA& 20A-492 & 2020 May 25 & 1090 & 6 & $<$0.018\\
 & --- & 22B-205 & 2022 Oct 7 & 1955 & 6 & $<$0.011\\
\hline
AT2018fyk & MeerKAT & SCI-20210212-YC-01 & 2021 May 8 & 973 & 1.36 & $<$0.060\\
& --- & SCI-20220822-YC-01 & 2023 Jan 25 & 1601 & 1.36 & $<$0.083\\
\hline
AT2018lna  & VLA& 21A-303 & 2021 Jun 20 & 906 & 6 & $<$0.018\\
 & --- & 22B-205 & 2022 Dec 8 & 1442 & 6 & $<$0.012\\
\hline
\enddata
\tablecomments{Upper limits quoted as  $3\sigma$.  This table only includes previously unpublished data for these TDEs; for published observations see Section \ref{sec:descriptions}.  We note that the uncertainties listed in this table are statistical only and do not include an expected $\approx 3-5\%$ systematic uncertainty in the overall flux density calibration, which is taken into account in the equipartition analysis (Section \ref{sec:equi}).}
\label{tab:obs}
\end{deluxetable}

\startlongtable
\begin{deluxetable}{llllllr}
\tablecolumns{7}
\tablecaption{All Other Radio Observations}
\tablehead{
Object&
Telescope&
Project Code/ &
Date of Observation&
$\delta t$ &  
Frequency  & 
Flux Density \\
&
& Source
& 
& 
(d)   & 
(GHz) & 
(mJy) 
}
\startdata
\hline
OGLE17aaj& ASKAP & VAST  & 2020 May 9 & 1224 & 1.42 & <0.3\\
& --- & --- & 2020 May 16 & 1230 & 1.42 & <0.3\\
& --- & --- & 2021 Apr 9 & 1391 & 1.42 & <0.3\\
& MeerKAT & --- & 2021 May 1 & 1581 & 1.36 & 0.193 $\pm$ 0.020\\
& ATCA & C3472 & 2022 Apr 9 & 1924 & 2.1 & <0.156\\
& --- & --- & --- & --- & 5.5 & <0.057\\
& --- & --- & --- & --- & 9 & <0.045\\
& --- & --- & --- & --- & 17 & <0.087\\
& --- & --- & --- & --- & 19 & <0.126\\
& MeerKAT & DDT-20220414-YC-01 & 2022 Apr 16 & 1931 & 0.82 & 0.290 $\pm$ 0.030\\
& --- & --- & --- & --- & 1.36 & 0.215 $\pm$ 0.023\\
& --- & SCI-20220822-YC-01 & 2022 Nov 24 & 2153 & 0.824 & 0.252 $\pm$ 0.05\\
& --- & --- & 2022 Dec 3 & 2161 & 1.36 & 0.249 $\pm$ 0.023\\
& --- & --- & 2023 Jan 5 & 2195 & 0.82 & 0.231 $\pm$ 0.025\\
\hline
AT2018bsi  & VLA& 21A-303 & 2021 Jun 20 & 1169 & 6 & $<$0.013\\
 & --- & 22B-205 & 2022 Dec 8 & 1705 & 6 & 0.037 $\pm$ 0.005\\
\hline
AT2018hyz & VLA & VLASS 3 & 2023 Feb 7 & 1578 & 3 & 16.838 $\pm$ 0.205\\
\hline
AT2020nov & VLA & 20A-372 & 2020 Oct 16 & 111 & 15 & 0.224 $\pm$ 0.009\\
 & --- & --- & 2021 Feb 10 & 228 & 6 & 0.376 $\pm$ 0.010\\
 & --- & --- & 2021 Feb 28 & 246 & 6 & 0.370 $\pm$ 0.015\\
 & VLA & 22B-205 & 2022 Nov 13 & 869 & 6 & 0.386 $\pm$ 0.019\\
\hline
AT2020mot & VLA & 22B-205 & 2022 Nov 18 & 887 & 6 & 0.086 $\pm$ 0.008\\
\hline
AT2020pj & VLA & VLASS 1 & 2020 Sept 19 & 261 & 3 & $<$.0.465\\
& --- & 22B-205 & 2022 Oct 28 & 1030 & 6 & 0.118 $\pm$ 0.006\\
& --- & --- & 2022 Dec 15 & 1078 & 6 & 0.093 $\pm$ 0.011\\
 & --- & VLASS 2 & 2023 Jan 24 & 1118 & 3 & $<$.0.454\\
 \hline
AT2020wey & VLA& 20A-372 & 2020 Oct 30 & 22 & 15 & $<$0.014\\
&--- & --- & 2021 Feb 13 & 128 & 6 & 0.0408$\pm$ 0.081\\
&--- & 22B-205 & 2022 Oct 7 & 729 & 6 & $<$0.018\\
\hline
\enddata
\tablecomments{Upper limits quoted as  $3\sigma$.  We note that the uncertainties listed in this table are statistical only and do not include an expected $\approx 3-5\%$ systematic uncertainty in the overall flux density calibration, which is taken into account in the equipartition analysis (Section \ref{sec:equi}).}
\label{tab:obs-not-used}
\end{deluxetable}

\bibliography{sample631}{}
\bibliographystyle{aasjournal}

%% This command is needed to show the entire author+affiliation list when
%% the collaboration and author truncation commands are used.  It has to
%% go at the end of the manuscript.
%\allauthors

%% Include this line if you are using the \added, \replaced, \deleted
%% commands to see a summary list of all changes at the end of the article.
%\listofchanges

\end{document}